# Populist Discourse and Entrepreneurship:
# The Role of Political Ideology and Institutions

Daniel L. Bennett
Department of Entrepreneurship
University of Louisville
College of Business W213
Louisville, KY40292
Email: BennettEcon@gmail.com
Tel: +001 502-852-7742
ORCID: 0000-0001-9911-3926

Christopher Boudreaux
Department of Economics
Florida Atlantic University
777 Glades Rd.
Boca Raton, FL 33431
Email: cboudreaux@fau.edu
Tel: +001 386-690-0630
ORCID: 0000-0002-7098-7184

Boris Nikolaev
Department of Entrepreneurship
Baylor University
One Bear Place #98011
Waco, TX 76798
Email: BorisNikolaev@gmail.com
Tel: +001 813-401-9756
ORCID: 0000-0001-6567-8494

**Abstract**

Using institutional economic theory as our guiding framework, we develop a model to describe how populist discourse by a nation's political leader influences entrepreneurship. We hypothesize that populist discourse reduces entrepreneurship by creating regime uncertainty concerning the future stability of the institutional environment, resulting in entrepreneurs anticipating higher future transaction costs. Our model highlights two important factors that moderate the relationship. First is the strength of political checks and balances, which we hypothesize weakens the negative relationship between populist discourse and entrepreneurship by providing entrepreneurs with greater confidence that the actions of a populist will be constrained. Second, the political ideology of the leader moderates the relationship between populist discourse and entrepreneurship. The anti-capitalistic rhetoric of left-wing populism will create greater regime uncertainty than right-wing populism, which is often accompanied by rhetoric critical of free trade and foreigners but also supportive of business interests. The effect of centrist populism, which is often accompanied by a mix of contradictory and often moderate ideas that make it difficult to discern future transaction costs, will have a weaker negative effect on entrepreneurship than either left-wing or right-wing populism. We empirically test our model using a *multi-level* design and a dataset comprised of more than 780,000 individuals in 33 countries over the period 2002-2016. Our analysis largely supports our theory regarding the moderating role of ideology. Still, surprisingly, our findings suggest that the negative effect of populism on entrepreneurship is greater in nations with stronger checks and balances.

**INTRODUCTION**

Populism is on the rise around the world —from the election of Donald Trump in the United States and the Brexit vote in Britain to the ascent of populist leaders in nations such as Brazil, Italy, Germany, India, Indonesia, and Poland, among others (Müller, 2016; Pappas, 2019). Given that many believe populism will remain one of the most influential political forces for the foreseeable future that has important ramifications for the institutional contexts that govern firm behavior (Dalio et al., 2017; Mudde, 2019), understanding the implications of populism for business is crucial. As Devinney and Hartwell (2020, p. 34) highlight, "the rise of populism in all of its varieties points to a weakness in our understanding of global and domestic institutions and their impact and relevance to international business." Although international business (IB) scholars widely acknowledge the importance of the traditional political structure (e.g., institutions) for business (e.g., Meyer & Peng, 2005; Peng et al., 2008), much IB research implicitly treats institutions as driverless vehicles devoid of the influence of politics and political actors such as populist leaders (Hartwell & Devinney, 2021). This presents, as Hartwell and Devinney (2021, p. 2) describe, a vast opportunity for IB scholars to understand how populism "may influence and impact firm behavior."

In this paper, we seize this opportunity by examining the relationship between populism and new venture creation from a comparative IB perspective (Terjesen et al., 2016). Specifically, we address the research question: *How does populist discourse by a nation's political leader influence individual decisions to pursue opportunity-motivated entrepreneurship (OME)?* Using institutional economic theory as our overarching framework (North, 1991; Williamson, 2000), we develop a model to explain how populist discourse by the nation's chief political executive reduces entrepreneurship. Populist discourse creates regime uncertainty by undermining confidence among entrepreneurs that a nation's institutions will continue to support market exchange and protect business and private property rights in the future, resulting in entrepreneurs anticipating a future increase in transaction costs and reducing their incentive to engage in new venture creation (Bylund & McCaffrey, 2017; Frølund, 2021). Our theoretical model also highlights two important boundary conditions that moderate the relationship between populist discourse and entrepreneurship, namely (1) the political checks and balances in place and (2) the ideology of the political leader.

Checks and balances serve as important institutional guardrails that constrain the actions of a populist leader (Stöckl & Rode, 2021), giving entrepreneurs more confidence that a populist will fail to succeed in fully implementing their agenda. Strong checks and balances, therefore, reduce the regime uncertainty associated with populist discourse such that entrepreneurs discount the possibility that transaction costs will rise substantially in the future, thereby weakening the negative effect of populist discourse on entrepreneurship. Because populism is a "thin-centered" ideology that often aligns with other thick ideologies (Mudde & Rovira Kaltwasser, 2018), the variety of populism will influence the magnitude of regime uncertainty and commensurate rise in transaction costs expected by entrepreneurs, thereby differentially affecting entrepreneurship. Specifically, the attachment of left-wing populism to socialism is often accompanied by harsh anti-capitalism rhetoric. At the same time, the attachment of right-wing populism to neoliberalism/nativism is often accompanied by rhetoric that is critical of free trade and foreigners but also supportive of business interests (Stöckl & Rode, 2021). Given these ideological differences, entrepreneurs will view left-wing populism as a greater threat to the pro-market institutional environment than right-wing populism. The former will exert a more adverse effect on entrepreneurship than the latter. Meanwhile, entrepreneurs will view the more moderate and sometimes contradictory views of centrist populists as less threatening than either left-wing or right-wing populism (Stanley, 2017). Our theory suggests that "who is the leader matters beyond ideology and institutional structure," for entrepreneurship, thereby partially addressing recent calls for IB scholars to extend institutional theory to better account for the role of politics and political actors (Devinney & Hartwell, 2020, p. 34; Hartwell & Devinney, 2021).

We assemble a unique dataset to empirically test our hypotheses using *multi-level* mixed-effects logit models. Specifically, we combine individual and national-level data from a variety of secondary sources, including the Global Entrepreneurship Monitor (GEM) micro survey, the Global Populism Dataset (GPD), and the Varieties of Democracy database (V-DEM), among others. Our final multi-level dataset consists of more than 780,000 individuals living in 33 nations with diverse institutional environments spanning the period 2002-2016. Importantly, we condition our models on a variety of individual and country-level factors, including the pro-market institutional environment and several measures of macro-economic uncertainty. This allows us to isolate the effect of regime uncertainty created by populist discourse from the effect of institutional quality and other potential sources of environmental uncertainty that may confound the relationship.

## THEORETICAL FRAMEWORK

Using institutional economic theory as our overarching framework, we develop the theoretical model depicted in Figure 1 to describe how populist discourse by a nation's political leader creates regime uncertainty that, by raising the anticipated future transaction costs facing entrepreneurs, discourages entrepreneurship. Our model also depicts how political checks and balances and the political ideology of the leader moderate the relationship between populist discourse and entrepreneurship. Next, we describe our conceptualization of populism as discourse by a nation's political leader. Following Devinney and Hartwell (2020), we take the existence of populism as a given and do not attempt to explain the drivers of the phenomenon.

[Figure 1]

**Populism**

Populism is a contested concept, or one that involves a considerable dispute about its proper use (Mudde, 2017), that has been used to describe a wide range of socio-political phenomena and contradictory ideologies (Mudde & Rovira Kaltwasser, 2017). As noted by Weyland (2001, p. 1), for instance, "a wide variety of governments, parties, movements, leaders, and policies have been labeled populist, and scholars have found populism to have radically divergent characteristics." Although the term populism has been applied very broadly, leading some scholars to refer to it as a "shifty" concept (Weyland, 2017), it plays a vital role in contemporary politics and is likely to continue to do so for the foreseeable future (Mudde, 2019; Müller, 2015). Despite the lack of a generally accepted definition of populism, several different conceptualizations of the phenomenon have emerged in the literature.

According to the Oxford Handbook of Populism (Rovira Kaltwasser et al., 2017), three distinct approaches have emerged in the political science literature: (1) ideational, (2) political-strategic, and (3) socio-cultural. An additional approach, often referred to as (4) economic, has been utilized in the economics literature (Absher et al., 2020; Acemoglu et al., 2013; Bittencourt, 2012; Dornbusch & Edwards, 1990). Table 1 summarizes the various approaches. While each approach focuses on a different aspect of the populist phenomenon and has both advantages and limitations, we follow an increasing number of scholars in adopting the ideation approach (Hawkins et al., 2018). This decision was based on three considerations.

[Table 1]

First, our study seeks to advance institutional economic theory beyond focusing on "faceless" institutional structures to better account for the role of political actors, who are the drivers of institutional and policy evolution, as determinants of entrepreneurial outcomes (Devinney & Hartwell, 2020; Hartwell & Devinney, 2021). Doing so requires a conceptualization of populism that captures the essence of the phenomenon, but is orthogonal to the actions of political leaders that are associated with institutional and policy changes. This is because the literature has established that institutions and economic policy are important antecedents of entrepreneurship (Bjornskov & Foss, 2016), and it is important, both theoretically and empirically, to demonstrate that populism has a direct effect on entrepreneurship that is distinct from the effect of a political leader's institutional and policy actions. This rules out the economic approach, which defines populism by myopic macroeconomic policy actions (Dornbusch & Edwards, 1990). It also rules out the political-strategic approach, which emphasizes the strategic and opportunistic actions of political actors to gain influence and sustain their authority because these actions are often embedded in actions intended to change the institutional environment (Weyland, 2017). Meanwhile, the ideation approach is based on a minimalist definition of populism, which we describe further below, that clearly distinguishes between populist and non-populist actors (e.g., elitists and pluralists) and allows for categorization of different types of populism, but is unrelated to any particular institutional or policy action of the political leader (Mudde, 2017).

Second, our study takes a comparative international entrepreneurship approach, which involves the "cross-national-border comparison of entrepreneurs, their behaviors, and the circumstances in which they are embedded" (Terjesen et al., 2016, p. 300). Accordingly, we need to conceptualize (and measure) populism in a manner that facilitates comparative international research. This necessitates a definition that can be applied consistently across both different geographical regions and time periods. The ideation approach has "travelability" (Mudde, 2017), meaning that it captures the essence of most political figures who are described as populist in different times and places (Mudde & Rovira Kaltwasser, 2018). Because it is characterized by distinguishability, categorizability, and travelability, the ideation approach is more suitable for empirical research and developing a broader comparative research agenda than other conceptual traditions (Hawkins et al., 2018; Mudde, 2017; Mudde & Rovira Kaltwasser, 2018).

Relatedly, our final consideration is a pragmatic one, namely data availability. Our study involves an empirical analysis to test our theoretical framework. This necessitates that we are able to observe comparable measures of populism over time for countries located in different regions of the world. The surge of interest in populism among scholars has been accompanied by attempts to develop new populism datasets using a variety of approaches. Still, most of the available datasets have limited geographical coverage and/or are based on conceptualizations that do not facilitate comparative international research. One exception, which we discuss in more detail in the data section below, is the Global Populism Dataset (Hawkins et al., 2019), which is based on the ideation approach and measures the extent of populist discourse by the chief political leader for a large number of countries located in different regions of the world. Next, we further describe the ideation approach.

The ideation approach provides a minimal definition of populism, depicting it as a "thin-centered" ideology or a set of ideas that consider society as divided between two homogenous and antagonistic camps, 'the people' and 'the corrupt elite'(Mudde & Rovira Kaltwasser, 2018). It also argues that politics are an "expression of the volonté générale (general will) of the people" (Mudde, 2004, p. 543). Populism has been conceived of in different ways, e.g., as a discourse (e.g., Stavrakakis & Katsambekis, 2014), ideology (Stanley, 2008), moral imagination (Müller, 2016), style (Moffitt, 2016), or worldview (Hawkins, 2010). Although ideation scholars disagree on the precise definition of the term, their differences are minor and "point to a fairly similar phenomena and populist forces," suggesting that research based on the ideation approach is complementary, cumulative, and fosters conversation between scholars from different fields (Mudde, 2017; Mudde & Rovira Kaltwasser, 2018, p. 1669),.

Within the ideation tradition, we adopt a discursive conception of populism, which focuses on a set of ideas expressed by political leaders in a Manichean discourse that makes a moral distinction between things that are good, as manifested by the will of the people and represented by the populist leader, and those that are evil, or those that subvert the will of the people as a result of a common enemy of conspiring elites (Hawkins, 2009). Populist leaders claim to represent "the people," which is not an inclusive or pluralistic group representative of all of the people living in a society, but rather an "empty signifier" that allows the populist to appeal to different constituent groups in order to create a shared identity and facilitate common cause (Mudde and Rovira Kaltwasser, 2017). As such, "the people" may exclude specific groups whose rights and interests in a liberal democracy are deemed as undermining the will of "the people" (Canovan, 1999).

Populist discourse is preoccupied with the discovery and identification of the enemy, based on a concept of morality, as this process helps negatively constitute "the people" (Hawkins, 2009; Mudde, 2017). Enemy lines can be drawn based on ethnic, native, or social status such that the enemy may include foreigners, minorities, the press, and/or the business and political class (Canovan, 1999; Müller, 2016). Furthermore, populists detest the political establishment and other "corrupt elites" that work against the general will of the people. The rights of the enemy groups are deemed expendable to realize this will. In other words, "the evil minority ceases to have legitimacy, citizenship, or possibly human rights because it has chosen to fight against the common good" (Hawkins, 2009, p. 1044).

To achieve the will of the people, some form of liberation or revolution is necessary to subvert the existing order that is suppressing it (Laclau, 2005). As Hawkins (2009, p. 1044) describes: "The old system has been taken over by the forces of evil and no longer serves the people. This conflict is not over particular policies or issues but institutions and the system. These must be reshaped or at least substantially modified; if not, the forces of evil will regroup and continue their oppression." Devinney and Hartwell (2020, p. 41) add, "populism is perpetuated…as a way to erode or change existing institutional structures from within, using the levers of democracy to forever alter its composition and shape." Former Venezuelan President Hugo Chávez, for example, famously declared that with his re-election, "another era will begin, another revolutionary era" (Hawkins, 2009, p. 1044).

Similar to rhetoric, populist discourse manifests itself in "distinct linguistic forms and content that have real political consequences" (Hawkins, 2009, p. 1045). Populist discourse combines elements of both ideology and rhetoric. Similar to ideology, populist discourse is based on a latent set of beliefs about how the world works. Although this discourse tends to compel political action among its believers, it typically "lacks significant exposition and…is usually low on policy specifics" (Hawkins, 2009, p. 1045). In this way, populism is a "thin-centered" ideology that, unlike thick ideologies, does not prescribe a "broad menu of solutions to major socio-political issues" (Freeden, 2003, p. 96). Rather, populism offers "few specific views on political-institutional or socio-economic issues" but is often attached to another host ideology (Mudde, 2017, p. 30).

Although the ideation approach to populism, and in particular the discursive approach, facilitates comparative international research and has become "dominant in the political science literature," social scientists in the positivist tradition have raised several concerns with it (Hawkins & Kaltwasser, 2017, p. 527). First is an epistemological concern regarding whether the concept can be measured and subjected to scientific hypothesis testing. As Hawkins and colleagues have demonstrated, measuring populist discourse is clearly feasible and comparable across countries and time (Hawkins, 2009; Hawkins et al., 2019; Hawkins & Rovira Kaltwasser, 2018). We discuss the measurement of populist discourse in more detail below in the data section. Second is an ontological concern of whether populist speech constitutes populism if it is not coupled with action. While Hawkins (2009, p. 1047) acknowledges that manifestations of populism cannot "exist without some material component… discourse is meaningless unless believed and shared by actual human beings," he argues that discourse is the "defining attribute of populism" and that "actions alone…are insufficient conditions for populism." Actions are ultimately

`populist' because of the meaning that is ascribed to them by their participants, not because of any objective quantity that adheres in them." A third concern is a real-world relevance. In other words, "if we accept discourse as a defining attribute of populism, does it matter"? As Hawkins (2009, p. 1047) notes, "If we can measure populist discourse and calculate its correlation with aspects of politics and economics that interest us, then we have shown that it matters. The question then becomes the more theoretically enriching one of how or why it matters." With these concerns in mind, our model importantly holds constant policy and institutional factors that are subject to populist leader manipulation in order to isolate the effects of their discourse from their actions. In doing so, we are able to examine whether, and under what boundary conditions, populist discourse affects entrepreneurs.[1]

**Institutional Economic Theory and Entrepreneurship**

Entrepreneurship is an experimental decision-making process that takes place in a market setting that is characterized by fundamental uncertainty, resource heterogeneity, and agents with cognitive and behavioral limitations. In the pursuit of profits, entrepreneurs combine heterogeneous assets within a firm to produce goods and services that they believe will satisfy consumers' wants. Anticipating future market conditions, entrepreneurs act upon their subjective assessments about resources, technological requirements, consumer preferences, and their expectations about the future profitability and growth of their firms. As such, new venture creation is a manifestation of a subjective entrepreneurial judgment process in which an entrepreneur takes action by acquiring and committing resources to production, based on their belief that a perceived business opportunity is economically sustainable (Foss et al., 2019; Foss & Klein, 2012). Such entrepreneurial judgments take place in a market context that is regulated and shaped by the quality (Foss et al., 2019) and stability (Young et al., 2018) of the institutional environment.

Institutions are "the humanly devised constraints that structure political, economic, and social interactions" that create order and (North, 1991, p. 97) reduce uncertainty in exchange (Knight, 1921). By doing so, institutions "play an essential role in ensuring the effective function of the market mechanism so that firms may engage in market transactions without incurring excessive transaction costs" (Young et al., 2018, p. 410). Because new venture creation requires long-run planning that is characterized by fundamental market uncertainty (McMullen & Shepherd, 2006), a strong institutional environment that supports "the voluntary exchange underpinning an effective market mechanism" (Meyer et al., 2009, p. 63) enables productive entrepreneurship by providing a stable structure that makes it "easier for decision-makers to anticipate the future, mitigating the effects of uncertainty" (Foss et al., 2019, p. 1207). Frølund (Frølund, 2021, p. 3) refers to this uncertainty-reducing characteristic of institutions as institutional clarity, which facilitates the entrepreneurial judgment process by allowing entrepreneurs to "choose whether to take action or not."

While it is widely acknowledged that pro-market institutions – which lower transaction costs by enabling "supply and demand-based exchanges, price clearing mechanisms…[and providing] rules, regulations, property rights protection, and contract dispute resolution mechanisms that reduce exchange hazards" (Banalieva et al., 2018, p. 860) – are an important antecedent to entrepreneurship (Bjornskov & Foss, 2016), entrepreneurship research has placed relatively little emphasis on uncertainty of the institutional environment itself (Bylund & McCaffrey, 2017; Frølund, 2021). Similarly, IB scholars widely acknowledge the importance of the institutional context for the international business phenomenon but have not yet incorporated the role of institutional uncertainty into their models and analyses (Hartwell & Devinney, 2021).

Institutional uncertainty is a particular source of perceived environmental uncertainty (Milliken, 1987) that exists when there is a perceived increase in the level of unpredictability of the institutional environment (Laine & Galkina, 2017). An entrepreneur assessing the long-run viability of a new venture idea will consider the current institutional environment as well as their subjective assessment of the future environment (Bylund & McCaffrey, 2017). When an entrepreneur perceives a credible threat to the existing institutional order, doubts "about the relevance of existing institutions" create an environment that can undermine their judgment process (Zahra, 2020, p. 174). This results in institutional uncertainty regarding the future stability of the institutions that entrepreneurs rely on to exercise judgment about their beliefs (Frølund, 2021), which precede entrepreneurial action (Foss & Klein, 2020). Institutional uncertainty, which makes it difficult to interpret the institutional environment, imposes transaction costs on entrepreneurs that undermine the entrepreneurial judgment process, potentially thwarting entrepreneurial action (Bylund & McCaffrey, 2017; Frølund, 2021).

Institutional uncertainty can be attributable to numerous sources and occur at different institutional levels (Bylund & McCaffrey, 2017). One particular type of institutional uncertainty that is highly relevant for our study is *regime uncertainty*, or the pervasive lack of confidence among investors and entrepreneurs in their ability to foresee the extent to which future government actions will alter their private-property rights and other market-supporting institutions that they rely on to reduce uncertainty (Bylund & McCaffrey, 2017; Higgs, 1997). Regime uncertainty arises when there is "mistrust of the people in charge of the institutions," leading to perceived instability and unpredictability about the future institutional environment (Frølund, 2021, p. 5). As a consequence, regime uncertainty significantly raises the transaction costs facing entrepreneurs. Next, we contend that populist discourse by a nation's political leader is a source of regime uncertainty that, by raising the transaction costs facing entrepreneurs, distorts the entrepreneurial judgment process and potentially undermines entrepreneurial action.

**Populism & Entrepreneurship**

Entrepreneurship thrives when entrepreneurs "trust that society holds values beneficial to business and property rights" (Bylund & McCaffrey, 2017, p. 471). Because populist discourse is preoccupied with calls for overhauling the institutional structure, which supposedly unjustly protects the interests of the conspiring elite, in order for the will of the people to truly prevail (Hawkins, 2009), the emergence of a populist regime can initiate the breakdown of this trust. This is because populist discourse from the political leader may signal that an ideological shift is underway, empowered by the will of the people, that no longer respects market-supporting institutional arrangements that encourage entrepreneurship by allowing for "profits to accrue to the entrepreneurs whose decisions created them" (Bylund & McCaffrey, 2017, p. 471).

Entrepreneurs may therefore perceive populist discourse as foreshadowing a shift in the institutional environment, serving as a credible threat to the institutional environment because they believe that the institutional structure is more malleable when a populist is empowered by the "will of the people" (Devinney & Hartwell, 2020). Indeed, Rode and Revuelta (2015) show that populist discourse by a nation's political leader is associated with a reduction in pro-market institutions. Thus, the emergence of a populist leader whose discourse signals a change in societal values may diminish confidence among entrepreneurs that institutions will continue to protect private property rights, reliably enforce contracts in an even-handed manner, and support market relationships (Cuervo-Cazurra et al., 2019). This creates regime uncertainty by undermining trust among entrepreneurs concerning the future stability of institutions (Bylund & McCaffrey, 2017). Such instability "makes predictions about the future conditions of…markets unreliable or altogether impossible" (Zahra, 2020, p. 172). Thus, the regime uncertainty caused by populist discourse significantly raises the transaction costs that entrepreneurs anticipate facing in the future (Hartwell & Devinney, 2021). Faced with these increased transaction costs, potential entrepreneurs may be dissuaded from or at least temporarily delay entrepreneurial action. By significantly raising anticipated transaction costs and distorting the entrepreneurial judgment process (Frølund, 2021), the regime uncertainty created by populist discourse may leave entrepreneurs "with no choice but to restrain their actions altogether or exit the market" (Bylund & McCaffrey, 2017, p. 472).

**H1**: *Populist discourse is negatively associated with entrepreneurship.*

**Populism, Checks and Balances & Entrepreneurship**

Populism hinders entrepreneurial action by creating regime uncertainty concerning the future stability of the pro-market institutional environment, thereby resulting in entrepreneurs anticipating a rise in transaction costs. However, this uncertainty only arises when entrepreneurs perceive that a populist's discourse concerning forthcoming institutional changes is credible (Banalieva et al., 2018). Because pro-market institutional changes are an outcome of the political process (Acemoglu et al., 2005; Cuervo-Cazurra et al., 2019), the feasibility for populist leaders to enact major institutional changes is more constrained in nations with stronger checks and balances (L. Diamond, 2021; Weyland, 2020), which encourage government accountability and limit the ability of public officials to rule by fiat and indiscriminately extract resources from individuals and organizations (Acemoglu, Robinson, et al., 2013; Keefer & Knack, 2007). In this way, checks and balances serve as important guardrails that provide entrepreneurs with greater confidence in the future stability of the institutional environment (Stöckl & Rode, 2021). Therefore, entrepreneurs residing in nations with stronger checks and balances are less likely to perceive populist discourse as a threat to the future integrity of the pro-market institutional environment because the power of the populist to pursue their agenda is perceived to be more constrained (Devinney & Hartwell, 2020). As such, the regime uncertainty created by populist discourse will be weaker in nations with stronger checks and balances such that entrepreneurs discount the risk of rising transaction costs. Therefore, strong checks and balances weaken the negative impact of populist discourse on entrepreneurship.

**H2:** *The strength of the checks and balances of the political system of a country weakens the negative effect of populist discourse on entrepreneurship.*

**Populism, Political Ideology & Entrepreneurship**

Although populism is a "thin-centered" ideology, it "rarely travels alone" and almost always appears attached to other "thick" ideologies (Rovira Kaltwasser et al., 2017, p. 17). By combining populism with other host ideologies, populists are able to explain the world and advance their political agenda in a context-specific manner. As such, there exist varieties of populism that are influenced largely by the host ideology to which they are attached (Devinney & Hartwell, 2020) and should, therefore, be studied in combination with these host ideologies (Mudde & Rovira Kaltwasser, 2018). Although the ideation approach treats populism as a "belief system of limited range" (Mudde & Rovira Kaltwasser, 2018, p. 1669) and suggests that populist forces can arise from anywhere along the political ideology spectrum (Hawkins et al., 2018), populism almost always appears attached to either a left-wing or right-wing ideology as a means to promote political objectives and reforms with broad appeal (Devinney & Hartwell, 2020; Hawkins & Rovira Kaltwasser, 2018; Mudde & Rovira Kaltwasser, 2018). While we propose that populist discourse will, in general, create regime uncertainty increasing anticipated transaction costs facing entrepreneurs, thereby reducing entrepreneurial action, we contend that the effect is moderated by the political ideology of the populist. Before describing how populism may differentially influence entrepreneurship depending on the host ideology, we first discuss how we conceptualize political ideology.

We follow Hinich and Munger (1992, p. 14) in adopting a neo-Downsian (Downs, 1957) definition of ideology, which places the construct along an abstract left-right dimension, as an "internally consistent set of propositions…that have implications for (a) what is ethically good, and therefore what is bad; (b) how society's resources should be distributed; and (c) where power appropriately resides." This conceptualization of ideology is based on the assumptions that political parties organize themselves around ideologies rather than particular policy positions and that voters use ideology as a heuristic to differentiate between candidates' positions on a "large, but undefined set of possible policies" (Kitschelt et al., 2009, p. 759). One of the major factors differentiating between left-wing and right-wing political ideology is a view towards the appropriate role of government vis á vis markets in the economy (Cuervo-Cazurra et al., 2019; Jahn, 2011), with left-wing parties tending to favor more redistribution and a greater scope of government involvement in the economy than right-wing parties, which tend to favor a more market-oriented approach (Bjørnskov & Potrafke, 2013; Bjørnskov & Rode, 2019; Pickering & Rockey, 2013; Potrafke, 2010). To a great extent, the instrument employed by the Democratic Accountability and Linkages Project (DALP) to rate political party ideologies along the left-right dimension, which is summarized in Table 2 and discussed further in the data section below, reflects this distinction (Kitschelt et al., 2009). Centrist political parties are those with more moderate and often mixed views towards policy that does not clearly align with either the left-wing or right-wing ideologies.

[Table 2]

Given differences between how left-wing and right-wing ideologies view the role of the government in the economy, entrepreneurial responses to a populist leader's public discourse will depend on the host ideology. We utilize a comparative framework to hypothesize about the potential differential effects of varieties of populism on entrepreneurship. We first discuss populist discourse by a centrist political leader as a baseline for comparison. Centrist populism is ideologically "hollowed out" with parties and leaders competing for office based on "claims to competence and moral probity rather than distinct policy platforms," characterized by appeals "to the people against allegedly corrupt and incompetent mainstream elite" that emphasize the need to reform institutions without a set of clear and consistent programmatic principles (Stanley, 2017, pp. 143–144). Because centrist populists do not have a clear ideological bent, their platform may contain a mixture of contradictory proposals that are more moderate in nature than the policy positions of left-wing or right-wing populist leaders. Nonetheless, the use of populist discourse by a centrist leader will make it difficult to discern the future institutional environment, creating regime uncertainty that results in entrepreneurs anticipating an increase in future transaction costs (Devinney & Hartwell, 2020; Stanley, 2017).

In practice, "right-wing populism mainly constitutes combinations of populism and neoliberalism and/or nationalism" (Mudde, 2017, p. 37). This implies that right-wing populism is coupled with a confluence of reform ideas that could be viewed as either favorable or unfavorable by entrepreneurs, resulting in uncertainty concerning the transaction costs facing entrepreneurs (Stöckl & Rode, 2021). Consistent with nativist ideology, right-wing populist leaders often portray foreigners and immigrants as the adversary of the people (Mudde & Kaltwasser, 2013) and use this as cover to advance greater restrictions on immigration and protectionist measures (Hauwaert & Kessel, 2018; Mudambi, 2018; Rodrik, 2018). During his 2016 U.S. Presidential campaign, for instance, Donald Trump strongly criticized previously enacted free trade agreements and immigration policies as detrimental to U.S. workers and businesses and was particularly critical of trade with China, stating that "[w]e can't continue to allow China to rape our country" (Diamond, 2016), we must "stand up to China on our terrible trade agreements and protect every last American job" (Blake, 2016). He also proposed building a southern border wall to stem the flow of immigrants from Latin America. Such protectionist institutional reforms, if enacted, would create restrictions on international labor mobility, imports, and foreign investment in the domestic economy (Chari & Gupta, 2008), creating regime uncertainty and increasing anticipated transaction costs facing entrepreneurs (Cuervo-Cazurra & Dau, 2009a; Devinney & Hartwell, 2020).

Neoliberalism, however, is often used to describe pro-market institutional reforms that reduce government intervention in the economy (e.g., less redistribution, lower taxes, privatization, deregulation) (Bjørnskov, 2015). Because such pro-market institutions reduce transaction costs (Foss et al., 2019), entrepreneurs will view the prospect of pro-market reforms by a right-wing populist positively, potentially mitigating some of the regime uncertainty created by harsh anti-free trade and anti-immigrant rhetoric. Although the nativist aspect of Trump's 2016 campaign likely created regime uncertainty, particularly among foreign and internationally-oriented entrepreneurs (Piper, 2019), it was also coupled with other pro-market institutional reform ideas that were likely perceived as beneficial for domestic entrepreneurs (Brandon, 2018). As Trump noted, "On taxes, we are going to massively cut tax rates for workers and small businesses…We're going to get rid of regulations…and we are going to make it easier for young Americans to get the credit they need to start a small business and pursue their dream" (Blake, 2016)."

Under the guise of greater inclusion and economic empowerment (Mudde & Kaltwasser, 2013), left-wing populists frequently criticize the market system and "rely on socialism to advance a definition of 'the pure people' that embraces the socioeconomic underdog" (Mudde & Rovira Kaltwasser, 2018, pp. 1669–1670) as a means to build and maintain support (Barro, 2017). In this way, the discourse of left-wing populists emphasizes income redistribution "coupled with a harshly anti-capitalist rhetoric" (Stöckl & Rode, 2021, p. 53). For instance, Evo Morales, former Bolivian President, suggested that "the fight between the rich and poor" is a battle between capitalism and socialism and "capitalism is the worst friend of humanity" (CNN, 2008). Similarly, Hugo Chávez famously described capitalism as the "way of the devil and exploitation…[noting that] only socialism can really create a genuine society" (Reuters Staff, 2013).

Such anti-market discourse by left-wing populists signals the forthcoming replacement of pro-market institutions with significantly greater government control over the allocation of economic resources (Hauwaert & Kessel, 2018). The contemporary experience of left-wing populism in Latin America – characterized by industry nationalizations, price controls, profligate spending, and excessive monetary expansion (Absher et al., 2020; Barro, 2017; Bittencourt, 2012; Edwards, 2010; Flores-Macías, 2010; Rodrik, 2018) – suggests that left-wing populist discourse indeed provided a credible signal of forthcoming replacements of pro-market institutions. In describing the effects of Chávez's left-wing populism in Venezuela, for instance, Oscar Garcia Mendoza (2014, 2015a, 2015b), former CEO of Venezolana de Crédito, Banco Universal, indicated that the "nation has been looted…almost all the legal system has been razed, and a volcanic deluge of laws, decrees, regulations, and other legal instruments was unleashed for the purpose of destroying everything [that is the private sector] and communizing the nation…All these legal instruments…hinder and prevent development or [entrepreneurial] activity." As such, left-wing populist discourse creates a culture that shuns entrepreneurship (Ireland et al., 2007) by generating considerable regime uncertainty concerning the future preservation of pro-market institutions (Stöckl & Rode, 2021), thereby raising anticipated future transaction costs.

While we contend that all varieties of populism will generate regime uncertainty, the ideological position of a populist leader will differentially influence the scope of regime uncertainty and the resultant change in future transaction costs created by populist discourse. As such, the ideology of a populist leader will moderate the relationship between populist discourse and entrepreneurship. Unlike right-wing and left-wing variants of populism, which are typically aligned with neoliberal/nativist and socialist ideologies, centrist populism is not attached to a particular host ideology and may contain a mixture of contradictory proposals that are more moderate than the radical policy positions of right-wing and left-wing populists. The more moderate policy views of a centrist populist will not create as much regime uncertainty as either a left-wing or right-wing populist such that the level of uncertainty, and hence anticipated transaction costs, will be lower when a centrist populist is in power than when either a left-wing or right-wing populist holds office. As such, populist discourse by a centrist leader will have less of a negative effect on entrepreneurship than populist discourse by either a left-wing or right-wing leader. While both right-wing and left-wing populist discourse will create regime uncertainty that results in entrepreneurs anticipating an increase in transaction costs, the harsh anti-market rhetoric that characterizes left-wing populist discourse will create significantly greater regime uncertainty than the mixed elements of nativism and neoliberalism that characterize right-wing populism, which may be viewed by many entrepreneurs as more business-friendly. As such, entrepreneurs will anticipate a greater increase in transaction costs when a left-wing populist is in office than when a right-wing populist holds power. The following hypotheses summarize our theoretical model regarding the moderating role of political ideology.

**H3a**: *Left-wing political leaders strengthen the negative impact of populist discourse on entrepreneurship relative to centrist political leaders.*

**H3b**: *Right-wing political leaders strengthen the negative impact of populist discourse on entrepreneurship relative to centrist political leaders.*

**H3c**: *Right-wing political leaders weaken the negative impact of populist discourse on entrepreneurship relative to left-wing political leaders.*

## DATA & METHODS

### Data

*Research Context.* We combined data from numerous secondary sources to create a unique multi-level dataset consisting of more than 780,000 individuals residing in a mix of 33 middle- and high-income nations, most of which are located in Europe or Latin America, over the period 2002-2016. The number of countries that we were able to include in our study is limited by data availability. Table 3 lists the countries and political regimes in our sample, including the populist scores of each leader.

[Table 3]

*Dependent variable.* GEM (2019) defines entrepreneurship as "Any attempt at new business or new venture creation, such as self-employment, a new business organization, or the expansion of an existing business, by an individual, a team of individuals, or an established business." We follow GEM and the judgment-based approach in treating new venture creation as our entrepreneurial action construct (Foss & Klein, 2012). Specifically, we focus on opportunity-motivated entrepreneurs (OME) who are "pulled" into entrepreneurship because they judge that the potential rewards (e.g., higher income, greater autonomy) of acting on a new venture idea exceed the costs and risks. We utilize the GEM survey data to measure OME using a dummy variable (i.e., 1 if an individual became an entrepreneur to take advantage of a business opportunity and 0 otherwise). Following Boudreaux et al. (2019), we exclude individuals from our sample who are involved in entrepreneurship out of necessity rather than lump them into the zero category.

*Independent variable.* We follow Hawkins (2009) in measuring populism as discourse by the chief political executive. Measuring political discourse at the elite level of the chief political executive is sensible because populism is typically associated with the leader responsible for the creation and consolidation of a respective populist movement (Rode & Revuelta, 2015). It is also practical because speeches by political leaders are widely available and can be analyzed and rated according to the degree of populist discourse that is used, with the results comparable across countries and time (Hawkins, 2009). Specifically, we utilize data from the Global Populism Dataset (Hawkins et al., 2019). GPD utilizes the holistic grading textual analysis method, a technique widely used for the development of large-scale standardized exams, to assess the level of populist discourse in the speeches of 215 chief political executives (i.e., president or prime minister) from 66 countries over the period 2000-2018. GPD covers 279 government terms and includes more than 1,000 speeches. Each speech was assessed by at least two trained native language speakers who rated it, in its original language, on a three-point scale.

A speech is assigned a score of 2 (i.e., extremely populis) if it "expresses all or nearly all of the elements of ideal populist discourse, and has few elements that would be considered non-populist." A speech is assigned a score of 1 (i.e., mixed discourse) if it "includes strong, clearly populist elements but either does not use them consistently or tempers them by including non-populist elements." Meanwhile, a speech is assigned a score of 0 (i.e., non-populist or pluralist) if it "uses few if any populist elements…[and] lacks some notion of a popular will" (Hawkins et al., 2019, p. 2). The data exhibit a strong degree of inter-rater reliability (Hawkins, 2009). Each speech rater, for each leader-term, assessed four speeches representing different categories: (i) campaign speech – usually the closing or announcement speech; (ii) ribbon-cutting speech –to a small, domestic audient to commemorate an event; (iii) international speech – to an audience of foreign nationals outside the country; and (iv) "famous" speech –a widely circulate one representing the leader at their best. Each speech met a certain standard of length (i.e., 1,000—3,000 words) and should have been the most recent available speech within each category at the time of coding. The final populist score for each chief political executive was calculated as the average of the speeches over a single term in office (Hawkins, 2009). The lack of available data for some of our other measures, including notably the GEM survey data, attenuated our final sample to 83 political leaders covering 33 countries over the period 2002-2016.

*Moderating variables.* We test two moderation hypotheses in our analysis. First, we test whether the ideology of the political leader's party moderates the relationship between populism and OME. We adopted the left-right indicator from DALP, as reported by Hawkins et al. (2019). DALP rates the ideology of political parties based on country expert country assessments of party policy positions on five generic policy issues (i.e., public spending on the disadvantaged; state role in governing the economy; public spending; national identity; and traditional authority, institutions, and customs) and country-specific policy issues that are well-known to generate inter-party divisions, as well as an overall placement along the left-right political spectrum (Kitschelt et al., 2009). The overall placement rating is on a 0-10 scale, but Hawkins et al. (2019) code as left-wing if the party is at least 0.5 standard deviations below the mean, right-wing if the party is at least 0.5 standard deviations above the mean, and centrist if the party is within 0.25 standard deviations of the mean. For cases falling within 0.25—0.5 standard deviations of the mean, Hawkins et al. (2019) adjudicate using either the Chapel Hill Expert Survey for European Parties or the Political Representation, Parties, and Presidents Survey for Latin American. We recoded the indicator as a set of dummy variables, namely left-wing, right-wing, and centrist.

Second, we test if political checks and balances moderate the effect of populism on OME using the government accountability index from the V-Dem version 10.1 dataset, which provides historical institutional measures for more than 200 nations. The index is designed to measure institutional constraints on the arbitrary use of political power by government officials and is comprised of three sources of institutional accountability: vertical accountability (i.e., accountability to the population via free and fair elections); horizontal accountability (i.e., checks and balances between branches of government); and diagonal accountability (i.e., impartial oversight of government by civil society and the media). The index is standardized and increasing in checks and balances (Coppedge et al., 2020).

*Control variables.* We control for a large number of individual and country-level characteristics. This includes, following Boudreaux et al. (2019), a set of individual demographic and economic characteristics that are anticipated to influence entrepreneurial behavior, including gender, age and age squared, educational attainment, and household income, and several socio-cognitive traits, including fear of failure, self-efficacy, and opportunity recognition. Data for all of the individual-level characteristics were obtained from the GEM dataset. We also control for numerous other country-levels factors to condition on the economic, cultural, and geopolitical context of the countries in our sample. This is done to minimize the possibility of omitted variable bias attributable to potential confounding factors (Wooldridge, 2010).

First, we control for the degree to which a nation's economic institutions and policies support market activity using the Fraser Institute's Economic Freedom of the World Index (Gwartney et al., 2020), a composite measure on a scale of 0-10, derived from more than 40 distinct variables, that is often used as a measure of pro-market institutions in the entrepreneurship (e.g., Bennett & Nikolaev, 2019) and international strategy (Cuervo-Cazurra & Dau, 2009b; Dau, 2012) literatures. Pro-market institutions provide protection of private property, transparent and reliable enforcement of contracts, stable monetary policy, and fewer market-distorting government interventions in the economy (Banalieva et al., 2018; Bennett et al., 2017; Cuervo-Cazurra et al., 2019), thereby fostering productive entrepreneurship by reducing uncertainty in exchange and lowering transaction costs (Foss et al., 2019). Indeed, a growing body of evidence shows that pro-market institutions encourage more entrepreneurial activity (Bennett, 2021; Bennett & Nikolaev, 2021a; Bjørnskov & Foss, 2013; Boudreaux et al., 2019; Boudreaux & Nikolaev, 2019; Bradley & Klein, 2016; Dau & Cuervo-Cazurra, 2014; Gohmann, 2012; McMullen et al., 2008; Nikolaev et al., 2018; Sobel, 2008). Pro-market institutions are a vital control in our model because our theory is that the discourse by a populist leader creates uncertainty about whether the institutional environment will continue to support market activity in the future. By controlling for pro-market institutions, we ensure that our estimates of the effect of populist discourse are not conflated with the effects of institutional changes.

Second, we have theorized that populist discourse adversely affects entrepreneurial action by generating regime uncertainty; however, it is possible that other sources of environmental uncertainty confound the relationship. Although we control for pro-market institutions and checks and balances to isolate the effects of populist discourse from the effects of populist actions that manifest in institutional changes, we nonetheless include several measures of uncertainty: Macroeconomic Uncertainty (i.e., the standard deviation of inflation rates), Macropolitical Uncertainty (i.e., the standard deviation of relative political extraction (Feng, 2001), and Trade Uncertainty (i.e., the standard deviation of tariff rates). Additionally, some populists manage to stay in office for long periods of time (e.g., Recept Tayyip Erdoğan has effectively been in power in Turkey since 2003) such that people become acclimated to the substance, style, and idiosyncrasies of their public discourse. As such, the effects of a leader's discourse on entrepreneurial judgments may dissipate the longer a populist's tenure in office. To account for this, we controlled for political leader tenure. Leader tenure also serves as an inverse measure of political change, which is often associated with political uncertainty (Dai & Zhang, 2019), because longer tenure is associated with less political change and vice versa.

Third, we control for two measures of market demand: the level of economic development (per capita GDP) and the population size (Boudreaux et al., 2019). Fourth, previous research suggests that informal institutions matter for entrepreneurship (Autio et al., 2013; Bennett & Nikolaev, 2021b; Stephan & Uhlaner, 2010) so we control for several measures of national culture, including the prevalence of major world religions, Hofstede's (2010) and Schwartz's (1994) cultural dimensions, social trust, and religion (Henley, 2017; Zelekha et al., 2014). Finally, we control for European Union (EU) membership, which may serve as an external mechanism for controlling institutional decay attributable to the actions of populist leaders. Table 4 provides summary statistics and a correlation matrix.

[Table 4]

**Methods**

Our model combines individual-level observations with country-level measures of populist discourse, political ideology, and checks and balances. Additionally, we control for pro-market institutions, several measures of macro-economic uncertainty, the level of economic development and size of the population, the cultural context, and a large number of individual-level factors. We estimate our models using hierarchical linear modeling. It is important to utilize a multi-level design because standard techniques (e.g., OLS) underestimate standard errors, significantly increasing the possibility of Type 1 errors when data are clustered (Hofmann et al., 2000). Specifically, we use a multi-level logit regression model to estimate the effects of country-level factors on the individual-level decision to become an OME. In multi-level (i.e., mixed-effects) models, random effects denote group-specific factors assumed to influence the dependent variable (Boudreaux et al., 2019), assuming that the groups are drawn randomly from a larger population. Consistent with the previous studies, we cluster at the country level (Autio et al., 2013; Peterson et al., 2012).

More specifically, we use mixed-effects logistic regression to estimate the influence of country-level political factors on the likelihood that individuals participate in OME. This model assumes unobserved country-specific effects are randomly distributed with a mean of zero and constant variance, $u_c \sim iid(0, \sigma_u^2)$, that are uncorrelated with our explanatory variables. We thus use a random-effects, generalized least square (GLS) algorithm that allows regression intercepts (also known as a random intercepts model) to vary across countries (Peterson et al., 2012). GLS estimates fixed parameters and maximum-likelihood estimates (MLE) of variance components, which permits standard errors to vary across group clusters and assigns greater weights to groups with more reliable level 1 estimates, providing greater influence in the level 2 regression (Hofmann et al., 2000).

$$OME_{ict} = \gamma_{00} + \beta_1 Populism_{ict} + \mu X'_{ict} + \nu Z_{ict} + \lambda y_t + u_c + e_{ic} \qquad (1)$$

Equation 1 depicts our baseline specification, where *i*, *c*, and *t* denote the individual, country, and year, respectively; *OME* denotes opportunity-motivated entrepreneurship; *Populism* represents populist discourse; $X_{ct}$ and $Z_{ict}$ are matrices of country-level and individual-level control variables; $y_t$ denotes year dummies; and $e_{ict}$ is an idiosyncratic error term. Because we estimate a random intercepts model, the intercept component, $\gamma_{00} = \gamma_0 + \gamma_c$, consists of the overall intercept, $\gamma_0$, and a country-specific (i.e., random) intercept, $\gamma_c$. $\beta_1$, which captures the effect of populist discourse on an individual's decision to become an OME, is our primary parameter of interest, and $\mu, \nu,$ and $\lambda$ are parameter vectors. To test our moderation hypotheses, we add interaction terms between the relevant moderators and populism to equation 1.

Because we use a random-effects model, which uses quasi-time-demeaned data at the country level, we are able to account for deviations in time-varying country-level factors. This is important because our theory is that populist discourse will influence entrepreneurial action by shaping their perceptions about the future institutional environment. As such, we need to empirically isolate this effect from that attributable to a populist's actual institutional changes. By controlling for pro-market institutions, checks and balances, several measures of macro uncertainty – all of which exhibit variation over time for at least some countries —and leader tenure, we account for populist actions that alter the institutional environment. Additionally, we condition on the cultural and economic context in order to minimize potential omitted variable bias attributable to confounding factors (Wooldridge, 2010).

Our first model, which includes all controls in addition to the populism variable, tests the hypothesis (i.e., H1) that populist discourse by a nation's leader negatively influences individual-level OME. We observe significant country-level variance (i.e., ICC = 2.09%), which supports our choice of a multilevel model over a simple logistic regression model. Next, we add our moderator variables as additional controls in model 2. To test our moderating hypotheses, we interact populism with checks and balances and political ideology in models 3 and 4. Consistent with previous models, we observe statistically significant country-level variances in each model and a reduction in the variance of each model's random intercept. Moreover, the LR test of $\rho = 0$ rejects the null hypothesis that the variance in the random intercept is not statistically different from zero, providing further support for the choice of multi-level methods over logistic regression (Boudreaux et al., 2019).

**EMPIRICAL RESULTS**

**Main Results**

We first highlight that we follow recent editorial guidelines on best practices in international business (Meyer et al., 2017) and entrepreneurship (Anderson et al., 2019) on reporting and discussing findings. Specifically, we adopt the following best practices: (i) present point estimates as odds ratios rather than coefficients for logit regressions; (ii) plot the marginal effects with 95 percent confidence intervals for moderating variables; (iii) discuss the effect size beyond statistical significance and omit asterisks in tables; and (iv) refer to exact p-values and confidence intervals when discussing findings.

Table 5 presents the results from our multi-level logistic regression model. Model 1 reports the influence of populism on the odds of becoming an OME. The odds ratio in model 1 (0.98) indicates that, consistent with H1, a one-unit increase in the intensity of populist discourse is associated with a 2 percent (1 – 0.98) decrease in the likelihood that an average individual becomes an OME (p = 0.503; ci = [0.925 1.039]). In model 2, we control for the ideology of the political leader and political checks and balances. Contrary to H1, the odds ratio (1.008) indicates that a one-unit increase in populist discourse is associated with a 0.8 percent increase in the likelihood of OME (p = 0.809; ci = [0.946 1.075]). While the result in model 1 is consistent with H1, the negative effect of populist discourse on OME dissipates when controlling for political leader ideology and checks and balances, although both estimates are estimated imprecisely. As such, our findings do not support H1.

[Table 5]

Model 3 includes an interaction between populism and checks and balances to test H2, which suggests that the negative effect of populist discourse on OME is mitigated in countries with strong checks and balances. Contrary to H2, we observe that populism has a positive effect on OME (odds ratio = 2.097; p = 0.000; ci = [1.517 2.899]) that is reduced in countries with stronger checks and balances (odds ratio = 0.580; p = 0.000; ci = [0.546 0.618]). Figure 2, which shows the average marginal effect of populist discourse on OME by the level of checks and balances, indicates that the positive effect of populism on OME steadily declines as the level of checks and balances rises, turning negative for countries with a checks and balances score above 1.5. Because the vast majority of individuals in our sample live in countries with checks and balances scores above 1.22 (1 standard deviation below mean), and more than three-fourths live in a nation with a checks and balances score above 1.5, our results suggest populism has a largely negative effect on OME that is exacerbated as political checks and balances increases.

[Figure 2]

Model 4 includes interaction terms between populism and both left-wing and right-wing ideologies. In this specification, centrist ideology is the omitted baseline ideology, and the direct effect of populist discourse represents centrist populism. This model allows us to formally test H3a and H3b by testing whether the effects of left-wing populism and right-wing populism on OME differ from the effects of centrist populism by examining the respective interaction terms. It also allows us to test H3c by testing the equality of the coefficients of the interactions between populism and left-wing and right-wing ideology. The effect of populism (odds=1.409; p = .005; ci = [1.103  1.799]) suggests that a one-unit increase in populist discourse by a centrist political leader is associated with a 40.9 percent increase in OME. Considering the interaction terms, we observe that the effect of populist discourse on OME is, on average, lower for both left-wing and right-wing leaders. Consistent with H3a, the populism-left-wing interaction term (odds=0.433; p = 0.000; ci = [0.386  0.486]) indicates that populist discourse by a left-wing leader, relative to centrist populism, reduces OME. The populism-right-wing interaction term (odds ratio = 0.975; p = 0.844; ci = [0.756  1.257]) also suggests that populist discourse by a right-wing leader reduces OME relative to centrist populism; however, the interaction term is estimated very imprecisely such that there is no discernible statistical difference in the effect of centrist and right-wing populism on OME. As such, our findings do not support H3b. However, consistent with H3c, the negative effect of left-wing populism on entrepreneurship is stronger than that of right-wing populism, as indicated by a p=0.000 from a test of the equality of coefficients testing the null hypothesis that the effect of populist discourse is different between left-wing and right-wing populists (i.e., $H_0$: 0.433 - 0.975 = 0).

To get a better understanding of how political ideology moderates the relationship between populist discourse and OME, we plot the marginal effects graphically. Figure 3 shows how populist discourse affects the average probability that an individual enters OME by the ideology of the political leader. As illustrated, more intense populist discourse reduces the probability of OME for left-wing leaders, but, surprisingly, it increases it for both centrist and right-wing leaders. The results suggest little moderation between right-wing and centrist leaders, with the major distinction between left-wing populists and the rest. The average probability an individual enters OME is higher for centrist leaders compared to left-wing and right-wing leaders, but it is only higher than right-wing leaders at low levels of populist discourse. Once populist discourse increases, there does not appear to be a statistically discernible difference between center leaders and right-wing leaders.[2] The graphical analysis provides additional support for H3a and H3c.

[Figure 3]

In addition to political ideology and checks and balances playing an important role in moderating the relationship between populist discourse and OME, we observe that both have discernible direct effects on OME. The odds ratios in model 4 for left-wing (1.034) and right-wing (0.818) indicate that the direct effect of having a political leader with a non-populist left-wing and right-wing ideology in office is a 3.4 (p = 0.277; ci = [0.970 1.102]) percent increase and a 18.2 (p = 0.000; ci = [0.775 0.863]) percent decrease in the likelihood of OME, relative to the baseline of having a centrist leader in office. Meanwhile, checks and balances enter as a positive determinant of OME. We observe checks and balances has a positive direct effect on OME (odds ratio = 1.026; p=0.000 ; ci =[0.783 0.899]).

**Additional Results**

We also performed a variety of post-hoc tests to explore the robustness of our findings, which are omitted for space but available upon request from the authors. Specifically, we conducted three different types of robustness tests (Meyer et al., 2017). First, we test the robustness of our results to alternative measures of our focal constructs. OME is our measure of entrepreneurship, but Baumol (1990) suggests that the allocation of entrepreneurial effort towards different types of activates is influenced by the institutional environment and previous research suggests that different types of entrepreneurship have different institutional antecedents (Chowdhury et al., 2019). Nikolaev et al. (2018), for instance, find that opportunity- and necessity-motivated entrepreneurship (NME) have different determinants. Similarly, Bowen and De Clercq (2008) and Estrin et al. (2013) find that the institutional context matters for high-growth aspiration entrepreneurship (HGE). It is conceivable, therefore, that the regime uncertainty created by populism may differentially influence entrepreneurial action, so we explored the robustness of our results using NME and HGE as alternative measures of entrepreneurial action. These results are qualitatively similar to our main results using OME.

Our measure of populist discourse reflects both pre- and post-election speeches by the political leader. Politicians may promise certain reforms that appeal to a mass of voters to garner support, even if such reforms are unlikely to be implemented in practice once elected (Grossman & Helpman, 2005). As such, entrepreneurs may not perceive a populist's campaign promises as a credible signal of forthcoming reforms such that pre-election discourse has little influence on their actions. We examine this possibility by re-estimating our baseline model separately using campaign speech discourse and the mean of the three other post-election speech types. The results for both measures are very similar to our main results, suggesting that populist discourse during a campaign can create regime uncertainty that deters entrepreneurial action, a finding that may reflect recent evidence that campaign promises are partially binding and increasingly fulfilled (Asako, 2015; Naurin et al., 2019).

Finally, we re-estimated our main models using several alternative estimators. First, we used country fixed-effects in lieu of random effects. By doing so, we account for all time-invariant country-level factors that may influence entrepreneurial action over time, allowing us to compare differences between individuals within the same country. The results are very similar to our main findings. Additionally, we explored whether the effects of populism on OME vary across countries by employing a random slopes model. These results suggest that populism has a positive effect on OME for a few countries in our sample (i.e., India, Peru, and the Netherlands), but on average, the effect is negative.

**DISCUSSION**

**Theoretical Contributions**

By exploring the relationship between populism and entrepreneurship, our study addresses the recent call by Audretsch and Moog (2020, p. 19) for entrepreneurship scholars to engage with theories and ideas from political science as well as Mudde and Rovira Kaltwasser's (2018) recent challenge to link the study of populism to academic fields outside of political science (Mudde & Rovira Kaltwasser, 2018). Specifically, we develop a theoretical model grounded in institutional economic theory to explain how populist discourse by a nation's chief political leader distorts entrepreneurial judgment by creating regime uncertainty that results in entrepreneurs anticipating a future increase in transaction costs, resulting in some entrepreneurs deciding to delay or completely forego entrepreneurial action (Bylund & McCaffrey, 2017; Frølund, 2021). Our model also highlights two important boundary conditions that moderate the relationship between populist discourse and entrepreneurship, namely checks and balances and the political ideology of the populist. By emphasizing "the role of political actors and personalities…and [showing] how who is involved in institutional evolution or function can influence…policy" (Devinney & Hartwell, 2020, p. 60), our study makes several important contributions to the literature.

While a growing body of the comparative international entrepreneurship literature suggests that institutions are an important country-level antecedent to entrepreneurship (Bjornskov & Foss, 2016; Terjesen et al., 2016), most of this research is focused on the importance of institutional quality (Dau & Cuervo-Cazurra, 2014; McMullen et al., 2008; Young et al., 2018). Meanwhile, very little attention has been paid to understanding the role of institutional instability or institutional dynamics. Our model suggests that populist discourse from the government leader can create a *perception* among entrepreneurs about forthcoming institutional changes (not necessarily actual institutional changes), leading to regime uncertainty that potentially undermines entrepreneurial action. Our framework suggests that both the quality and *perceived* stability of institutions are important for encouraging productive entrepreneurship (Baumol, 1990; Bowen & De Clercq, 2008). The idea that perceived institutional stability matters for entrepreneurship is consistent with the view that political economy perceptions serve as a "mechanism that filters the impact of objective conditions on individual-level processes" (Begley et al., 2005, p. 36). This insight suggests that future IB research should treat strong political institutions and political instability as distinct constructs and attempt to disentangle their effects (Hartwell & Devinney, 2021). It also complements calls from IB scholars to adopt a more dynamic view of the institutional environment (Banalieva et al., 2018; Griffith, 2010; Szyliowicz & Galvin, 2010), one that provides a new perspective on how the interaction between political instability and institutional reform influences national ecosystem development (Allard et al., 2012; Reuber et al., 2018).

Next, we theorize that the effect of populist discourse on entrepreneurship is conditional on the strength of checks and balances. In this way, we address recent calls from IB scholars to understand better the dynamics between the institutional and political environments (Banalieva et al., 2018; Cuervo-Cazurra et al., 2019; Devinney & Hartwell, 2020). Our findings, which contradict our theoretical prediction, suggest that the regime uncertainty created by populist discourse, and perceived institutional stability created by political turbulence more generally (Zahra, 2020), may be relatively innocuous in nations with weak checks and balances. Meanwhile, the negative effect of populist discourse on entrepreneurship is larger in nations with stronger checks and balances. This finding calls into question the "effectiveness of existing political institutions in preventing populism from having a vector to power" and may reflect the perception among entrepreneurs that populism is a threat to liberal democracy because populist leaders are adept at overcoming institutional hurdles (Devinney & Hartwell, 2020, p. 42; L. Diamond, 2021; Kaufman & Haggard, 2019; Pappas, 2019; Rummens, 2017; Weyland, 2020). This important boundary condition provides a "micro-focused understanding of institutions" (Devinney & Hartwell, 2020, p. 60) that may be useful for gaining deeper insights into the strategies employed by entrepreneurs in weak institutional environments (Baker et al., 2005; De Clercq et al., 2010; Khanna & Rivkin, 2001) during periods of political uncertainty (Cuervo-Cazurra, 2016; Tallman, 1988; Witt & Lewin, 2007).

Third, our study provides insights into how government ideology influences entrepreneurial behavior. Individuals living in countries whose leader has either left-wing or right-wing ideology are more likely to become an OME relative to those living under a leader with a centrist ideology. Given ideological differences concerning business and market intervention, this result is interesting and merits further research to understand better the mechanisms through which government ideology, and changes in ideology, encourage and/or discourage entrepreneurial action (Duran et al., 2017). More importantly, by modeling the interaction between the political leader and their political ideology, we examine the heterogeneous effects of different varieties of populism on entrepreneurial action (Devinney & Hartwell, 2020). We observe that ideology is an important moderator of the relationship between populist discourse and entrepreneurship. Relative to centrist leaders, populist discourse by left-wing political leaders reduces the likelihood that an individual enters entrepreneurship, but it is associated with a modest increase in OME for right-wing populists. These results are consistent with a recent study by Stöki and Rode (2021), who found that financial investors assign higher risk assessments to left-wing than right-wing populism, and affirm Devinney and Hartwell's (2020, p. 34) assertion that "who is the [political] leader matters beyond ideology and the institutional structure." This suggests that IB scholars should incorporate these important, yet often omitted, factors in their models and analyses.

**Practical Implications**

By some measures, populism is now at its highest level since the 1930s, making it one of the most influential political forces for the foreseeable future (Dalio et al., 2017). The number of populist leaders in power, for example, has increased five-fold since the 1990s, including not only countries in Latin America and Eastern Europe, where it has been most rampant, but also in Asia and Western Europe (Kyle & Gultchin, 2018). Mudde (2019, p. 1) suggests that populism is becoming "the concept that defines our age." Hartwell and Devinney (2021, p. 1) add that the "continuing anti-liberal stance of populist governments…have radically increased political and economic policy uncertainty globally and have the ability to influence the global environment for decades" to come. Given that populism continues to gain momentum and is likely to remain a key aspect of the political environment for the foreseeable future, our findings have several important implications for both policy-makers and entrepreneurs.

Policy-makers around the world are increasingly focused on encouraging entrepreneurship as a means to facilitate economic development, broadly construed. Our findings suggest that populism may be detrimental to such efforts, creating regime uncertainty that may undermine the ability of entrepreneurs to navigate markets. Contrary to the prediction of our model, our findings highlight that this is particularly true for countries with strong checks and balances. Such countries are typically characterized by independent legislatures and judiciaries intended to serve as checks on the arbitrary exercise of power by the executive. Legislators in countries that elect a populist leader can counteract some of the regime uncertainty by publicly speaking out against the damaging populist discourse of the executive, as well as opposing radical institutional and policy changes proposed by the populist leader. Judges can also play an important role in mitigating the pernicious effects of populism by contesting unconstitutional power grabs that undermine the rule of law.

What are the implications of our findings for entrepreneurs? As we have demonstrated, populist discourse creates regime uncertainty that, under certain conditions, reduces domestic entrepreneurial action. As such, entrepreneurs should keep abreast of the political conditions in their country and be concerned when politicians start espousing populist rhetoric, as this may provide a signal of forthcoming institutional changes that undermine their business and property rights. Entrepreneurs should be especially vigilant of left-wing populism, which tends to be attached to a socialist ideology, as our findings suggest it is particularly harmful to entrepreneurship. This may have been what alarmed Home Depot co-founder Bernie Marcus when he referred to U.S. Senator Bernie Sanders, a left-wing populist ( Hawkins & Rovira Kaltwasser, 2018) leading the Democratic Party presidential primary race at the time, as "the enemy of every entrepreneur that's ever going to be born in the country and has been born in the past" (Fox Business, 2019). Entrepreneurs should also be on the lookout for early signs of populist movements, including the development of populist attitudes among the general public that may suggest a demand for a populist leader (Hauwaert & Kessel, 2018). In addition to remaining aware of potential populist movements in their country, entrepreneurs concerned about the future impact of such forces may want to consider engaging in institutional entrepreneurship by partnering with other concerned business and community leaders to counteract these trends  (Aldrich, 2011; Bylund & McCaffrey, 2017).

In extreme circumstances, entrepreneurs may also want to consider escaping the regime uncertainty wrought by populism by relocating to a country with a more stable institutional environment (Cuervo-Cazurra, 2016; Witt & Lewin, 2007). Alternatively, entrepreneurs may find success in embracing populism and impending institutional change by investing in the development of their political capabilities (Holburn & Zelner, 2010). By aligning with the ideological position of a populist regime (Duran et al., 2017), an entrepreneur may gain access to new resources, learn how to navigate a rapidly changing institutional environment, and gain a competitive advantage in the market (Cuervo-Cazurra & Genc, 2008; Khoury et al., 2014). While such a political strategy may be economically beneficial for those entrepreneurs who manage to gain favor with a populist regime (Hartwell & Devinney, 2021), Klein et al. (2021) suggest that such allocations of entrepreneurial effort constitute unproductive acts of cronyism that, if left unchecked, discourage innovation and economic growth, a perspective supported by several macroeconomic studies on the growth consequences of populism (Absher et al., 2020; Bittencourt, 2012; Grier & Maynard, 2016).

**Limitations & Future Research Directions**

Like all studies, ours has some limitations that provide guidance for future research. First, the number of countries with full data is limited, so our country sample size is small. Furthermore, our sample consists of middle- and high-income democratic nations, many of which are located in Europe and Latin America. As such, our theoretical framework and empirical findings may not be generalizable to non-democratic contexts or other regions of the world. As additional data become available, it would be useful to revisit the relationship between populist discourse and entrepreneurial action to determine if our results are robust to a larger sample and/or non-democratic political contexts.

Because populism, as conceptualized by the ideation approach, is a democratic phenomenon, future research examining the effects of populism on entrepreneurship in non-democratic contexts will need to adopt an alternative conception of populism (Devinney & Hartwell, 2020). As such, a different framework and dataset will be needed to examine how different types of populism affect entrepreneurship, but such investigations will need to be more regionally focused since these approaches are not as well-suited for comparative international research (Mudde & Rovira Kaltwasser, 2018). Several alternative populism datasets, including the Timbro Authoritarian Populism Index (2019) for European political parties and several Latin American populism indices (Cachanosky & Padilla, 2019; Sáenz de Viteri & Bjørnskov, 2018), may facilitate such research.

Second, we interpret our findings as suggestive that populism is negatively associated with productive entrepreneurial action. However, regime changes such as those created by the emergence of a populist leader may also result in a reallocation of resources to entrepreneurs ideologically aligned with the new regime (Stöckl & Rode, 2021). It may also encourage some entrepreneurs to seek economic rents through the political process, resulting in a redirection of entrepreneurial effort towards unproductive, or even destructive, activities (Baumol, 1990; Boudreaux et al., 2017; Chowdhury et al., 2019; Sobel, 2008). As such, future research that examines entrepreneurial and firm success in populist regimes would help us better understand these mechanisms and the strategies that entrepreneurs employ in response to the regime uncertainty created by populist regimes (Cuervo-Cazurra & Genc, 2008; Khanna et al., 2005; Khanna & Palepu, 1997; Pinkham & Peng, 2017; Witt & Lewin, 2007), as well as the potential economic consequences of their actions (Frølund, 2021). Bylund and McCaffrey's (2017) institutional hierarchy misalignment framework may be useful to help understand entrepreneurial responses to populism in such contexts.

Next, our study is focused on the effect of home country populism on domestic entrepreneurship, but populism may have differential effects on different types of international entrepreneurship. Previous IB research, for instance, demonstrates the importance of the institutional environment for various types of foreign market entry, such as direct export (Aparicio et al., 2021; Terjesen & Hessels, 2009), foreign direct investment (Holburn & Zelner, 2010; Witt & Lewin, 2007), born-global ventures (Coeurderoy & Murray, 2008; Fan & Phan, 2007), and international joint ventures (Pinkham & Peng, 2017). Future research should explore the effects of populism on different types of international entrepreneurship. Such research could be done from the perspective of the home or the host country (Banalieva et al., 2018;

Cuervo-Cazurra, 2006) or even their cross-national distance (Berry et al., 2010; Cuervo-Cazurra & Genc, 2011).

Finally, while we control for a large number of individual and country-level factors, it is possible that there are unobservable confounding factors that bias our results. For instance, the regime uncertainty created by populist discourse may be less severe in nations with a long history of populism. Relatedly, our model and results suggest that the regime uncertainty created by populist discourse reduces entrepreneurial action; however, we only account for the short-run effects of electing a populist leader. Entrepreneurs may temporarily delay action while a populist is in office or until their discourse changes. For example, Robert Fico's discourse was much less populist during his second term as Slovakia's Prime Minister. Both of Fico's terms are included in our sample, so our analysis accounts for this change. Our dataset does not allow for us to control for the cumulative or long-run effects of populism, although we do to a limited extent account for these effects by controlling for leader tenure. As more historical data becomes available, future research can explore further the long-run term effects of populism.

## CONCLUSION

Although interest in populism has increased exponentially in recent years, there has been very little research on how such changes in the political landscape have affected businesses (Devinney & Hartwell, 2020; Mudambi, 2018). To the best of our knowledge, we are the first to examine how populism affects entrepreneurship. Drawing on institutional economic theory, we develop a model to depict how populist discourse by the political leader reduces entrepreneurship by generating regime uncertainty that distorts entrepreneurial judgment by undermining confidence in the stability of institutions and predictability of transaction costs. However, the regime uncertainty created by populist discourse is context-dependent. The ideological position of the political leader and the strength of checks and balances serve as important boundary conditions. Specifically, we find that left-wing populist discourse is much more damaging for entrepreneurship than right-wing or centrist populism discourse, and populist discourse is more harmful in nations with stronger checks and balances.

**Table 1. Summary of Approaches to Populism**

| Approach | Definition | Distinguishable | Categorizable | Travelable | Versatile |
|---|---|---|---|---|---|
| Economic | Myopic economic policies intended to appeal to the weaker classes that "emphasizes economic growth and income redistribution and deemphasizes the risks of inflation and deficit finance, external constraints and the reaction of economic agents to aggressive non-market policies" (Dornbusch & Edwards, 1990, p. 247; Rode & Revuelta, 2015) | No <br><br> "points to alleged consequences of populism but does not provide clear criteria for conceptualizing populism as such" (Rovira Kaltwasser et al., 2017, p. 14) | No <br><br> limits populism to "left-wing or inclusionary forms…[and] cannot grasp rightist or exclusionary expressions of populism that predominant in the various places of the world today" (Rovira Kaltwasser et al., 2017, p. 14) | No <br><br> "places redistributive objectives at the center of its definition, thus focusing exclusively on left-wing populist regimes in Latin America" (Rode & Revuelta, 2015, p. 76) | No |
| Ideational | A "thin-centered" ideology, often attached to a "thick" ideology that "considers society to be ultimately separated into two homogenous and antagonistic groups: `the pure people' versus `the corrupt elite', and which argues that politics should be an expression of the volonté générale (general will) of the people" (Mudde, 2004, p. 543, 2017) | Yes | Yes | Yes | Yes |
| Political-Strategic | "A political strategy through which a personalistic leader seeks or exercises government power based on direct, unmediated, uninstitutionalized support from large number of mostly unorganized followers" (Weyland, 2001, p. 14) | Yes | No <br><br> Chameleon-like: "constantly changes colors" (Weyland, 2017, p. 48) <br><br> Lacks "clear borderlines…there can be partial and mixed types that fixed conceptual categories do not fully capture" (Weyland, 2017, p. 65) | Yes | No <br><br> "essence of populism…revolves around top-down leadership" (Weyland, 2017, p. 53) |

**Table 1. Summary of Approaches to Populism**

| Approach | Definition | Distinguishable | Categorizable | Travelable | Versatile |
|---|---|---|---|---|---|
| Socio-Cultural | "a particular form of political relationship between political leaders and a social basis, one established and articulated through 'low' appeals which resonate and receive positive receptions within particular sectors of society for socio-cultural historical reasons" (Ostiguy, 2017, p. 73) | Yes | Yes | Yes | Yes |

*Notes: Distinguishable refers to clear boundaries established that delineate between populist and non-populist actors. Categorizable refers to feasibility of constructing logical taxonomies of populist actors. Travelable refers to comparability of populists across regions, nations, and time. Versatile refers to the ability to analyze populism at different levels (Mudde, 2017).*

**Table 2. Political Ideology Policy Issue Positions**

| Policy Issue | Left-Wing Ideology | Right-Wing Ideology |
|---|---|---|
| Social spending on the disadvantaged | Advocates extensive social spending redistributing income to benefit the less well-off in society | Opposes extensive social spending redistributing income to benefit the less well-off in societyv |
| State role in governing the economy | Supports a major role for the state in regulating private economic activity to achieve social goals, in directing development, and/or maintaining control over key services | Advocates a minimal role for the state in governing or directing economic activity or development |
| Public spending | Supports extensive public provision of benefits such as earnings-related pension benefits, comprehensive national health care, and basic primary and secondary schools for everyone | Opposes an extensive state role in providing such benefits and believes that such things as health insurance, pension, and schooling should be privately provided or that participation in public social insurance programs should be voluntary |
| National identity | Advocates toleration and social and political equality for minority ethnic, linguistic, and racial groups and opposes state policies that require the assimilation of such groups to the majority national culture | Believes that the defense and promotion of the majority national identity and culture at the expense of minority representation are important goals |
| Traditional authority, institutions, and customs | Advocates full individual freedom from state interferences into any issues related to religion, marriage, sexuality, occupation, family life, and social conduct in general. | Advocates state-enforced compliance of individuals with traditional authorities and values on issues related to religion, marriage, sexuality, occupation, family life, and social conduct in general. |

*Notes: Table adapted from Democratic Accountability and Linkages Project codebook.*

**Table 3. List of Leaders, Country, Ideology, and Party in the Sample**

| Leader | Country | Years | Ideology | Populism | Party |
|---|---|---|---|---|---|
| Eduardo Duhalde | Argentina | 2002-2003 | Center | 0.53 | Justicialist Party |
| Cristina Fernández | Argentina | 2007-2015 | Left | 0 | Front for Victory |
| Mauricio Macri | Argentina | 2015-2019 | Right | 0; 0.04 | Republican Proposal |
| Wolfgang Schüssel | Austria | 2000-2007 | Right | 0 | Austrian People's Party |
| Alfred Gusenbauer | Austria | 2007-2008 | Left | 0.288 | Social Democratic Party of Austria |
| Werner Faymann | Austria | 2008-2016 | Left | 0.08; 0.10 | Social Democratic Party of Austria |
| Sebastian Kurz | Austria | 2017-2019 | Left | 0.28 | Austrian People's Party |
| Fernando Henrique | Brazil | 1995-2002 | Center | 0.00 | Brazilian Social Democratic Party |
| Luiz Inácio Lula da Silva | Brazil | 2003-2010 | Left | 0.25 | Workers' Party |
| Dilma Rousseff | Brazil | 2011-2016 | Left | 0; 0.34 | Workers' Party |
| Michel Temer | Brazil | 2016-2018 | Center | 0.34 | Brazilian Democratic Movement Party |
| Rosen Plevneliev | Bulgaria | 2012-2017 | Right | 0.16 | GERB |
| Stephen Harper | Canada | 2006-2015 | Right | 0.23 | Conservative Party |
| Justin Trudeau | Canada | 2015-current | Center | 0.11; 0.23 | Liberal Party |
| Ricardo Lagos | Chile | 2000-2006 | Left | 0.08 | Party for Democracy |
| Michelle Bachelet | Chile | 2006-2010[a] | Left | 0.21 | Socialist Party of Chile |
| Sebastián Piñera | Chile | 2010-2014 | Right | 0 | National Renewal |
| Juan Manuel Santos | Colombia | 2010-2018 | Center | 0; 0.06 | Social National Unity Party |
| Ivica Račan | Croatia | 2000-2003 | Left | 0 | Social Democratic Party of Croatia |
| Ivo Sanader | Croatia | 2003-2009 | Right | 0; 0.5 | Croatian Democratic Union |
| Zoran Milanovic | Croatia | 2011-2016 | Left | 0.18; 0.50 | Social Democratic Party of Croatia |
| Kolinda Grabar-Kitarović | Croatia | 2015-2020 | Left | 0.18 | Croatian Democratic Union |
| Mirek Topolanek | Czech | 2006-2009 | Right | 0.63 | Civic Democratic Party |
| Petr Necas | Czech | 2010-2013 | Right | 0.15; 1.0 | Civic Democratic Party |
| Andrus Ansip | Estonia | 2005-2014 | Right | 0 | Estonian Reform Party |
| Tarja Halonen | Finland | 2000-2012 | Right | 0 | Social Democratic Party of Finland |
| Jacques Chirac | France | 2002-2006 | Right | 0.05 | Union for a Popular Movement |
| Nicolas Sarkozy | France | 2007-2011 | Right | 0.20 | Union for a Popular Movement |
| Francois Hollande | France | 2012-2016 | Left | 0.14 | Socialist Party |
| Emmanuel Macron | France | 2017-current | Center | 0.14 | Forward Republic |
| Antonis Samaras | Greece | 2012-2015 | Right | 0.35 | New Democracy |
| Alexis Tspiras | Greece | 2015-2019 | Left | 0.25 | Coalition of the Radical Left |
| Ferenc Gyurcsany | Hungary | 2004-2009 | Left | 0; 0.38 | Hungarian Socialist Party |
| Viktor Orban | Hungary | 2010-current | Right | 0.38; 0.83; 0.88 | The Fidesz - Hungarian Civic Alliance |
| Atal Bihari Vajpayee | India | 1998-2004 | Right | 0.04 | Bharatiya Janata Party |
| Manmohan Singh | India | 2004-2014 | Center | 0 | Indian National Congress |
| Narendra Modi | India | 2014-current | Right | 0.55 | Bharatiya Janata Party |
| Mahmoud Ahmadinejad | Iran | 2005-2013 | Right | 1.17 | Alliance of Builders of Islamic Iran |
| Berthie Ahern | Ireland | 1997-2008 | Center | 0.04 | Fianna Fáil |
| Brian Cowen | Ireland | 2008-2011 | Center | 0.03 | Fianna Fáil |
| Enda Kenny | Ireland | 2011-2017 | Right | 0.10 | Fine Gael |
| Romano Prodi | Italy | 2006-2008 | Left | 0.08 | Democratic Party |
| Silvio Berlusconi | Italy | 2008-2011 | Right | 0.75; 0.88 | Forward Italy |
| Matteo Renzi | Italy | 2014-2016 | Left | 0.04 | Democratic Party |
| Yasuo Fukuda | Japan | 2007-2008 | Right | 0.25 | Liberal Democratic Party |
| Taro Aso | Japan | 2008-2009 | Right | 0.25 | Liberal Democratic Party |
| Yukio Hatoyama | Japan | 2009-2010 | Right | 0.25 | Democratic Party of Japan |
| Naoto Kan | Japan | 2010-2011 | Right | 0.25 | Democratic Party of Japan |
| Yoshihiko Noda | Japan | 2011-2012 | Right | 0.25 | Democratic Party of Japan |
| Shinzo Abe | Japan | 2012-2020 | Center | 0; 0.05; 0.75 | Liberal Democratic Party |
| Aigars Kalvitis | Latvia | 2004-2007 | Right | 0.5 | The People's Party |
| Laimdota Straujuma | Latvia | 2014-2016 | Right | 0 | Unity |

| Leader | Country | Years | Ideology | Populism | Party |
|---|---|---|---|---|---|
| Valdis Dombrovskis | Latvia | 2009-2014 | Right | 0 | The New Era Party / Unity |
| Māris Kučinskis | Latvia | 2016-2019 | Right | 0 | Liepāja Party |
| Vicente Fox | Mexico | 2000-2006 | Right | 0.25 | National Action Party |
| Enrique Peña Nieto | Mexico | 2012-2018 | Right | 0 | Constitutionalist Liberal Party |
| Jens Stoltenberg | Norway | 2005-2013 | Left | 0.17; 0.21 | Labour Party |
| Erna Solberg | Norway | 2013-current | Right | 0.09; 0.21 | Conservative Party |
| Alejandro Toledo | Peru | 2001-2006 | Center | 0.33 | Possible Peru |
| Ollanta Humala | Peru | 2011-2016 | Left | 0.50 | Peruvian Nationalist Party |
| Leszek Miller | Poland | 2001-2004 | Center | 0.21 | Democratic Left Alliance |
| Donald Tusk | Poland | 2007-2014 | Right | 0.16 | Civic Platform |
| Beate Szydlo | Poland | 2015-2017 | Right | 0.16 | Law and Justice |
| Vladimir Putin | Russia | 2008-current | Center | 0; 0.03; 0.05; 0.5 | United Russia |
| Robert Fico | Slovakia | 2012-2018 | Left | 0.06; 0.10; 0.75 | Direction - Social Democracy |
| Janez Jansa | Slovenia | 2004-2008 | Right | 0.75 | Slovenian Democratic Party |
| Borut Pahor | Slovenia | 2008-2012 | Left | 0 | Social Democrats Party |
| José María Aznar | Spain | 1996-2004 | Right | 0.05 | The People's Party |
| José Luis Rodríguez | Spain | 2004-2011 | Left | 0; 0.36 | Spanish Socialist Worker's Party |
| Mariano Rajoy | Spain | 2011-2018 | Right | 0.01; 0.36 | The People's Party |
| Göran Persson | Sweden | 1996-2006 | Left | 0.00 | Social Democratic Party |
| Fredrik Reinfeldt | Sweden | 2006-2014 | Right | 0.00 | Moderate Party |
| Stefan Löfven | Sweden | 2014-current | Left | 0.10 | Social Democratic Party |
| Recep Tayyip Erdogan | Turkey | 2014-current | Right | 0.13; 0.88; 1.47 | Justice and Development Party |
| Tony Blair | UK | 1997-2007 | Left | 0.10; 0.13 | Labour Party |
| Gordon Brown | UK | 2007-2010 | Left | 0.08 | Labour Party |
| David Cameron | UK | 2010-2016 | Right | 0.01 | Conservative Party |
| George W. Bush | United States | 2000-2008 | Right | 0.19; 0.21 | Republican Party |
| Barack Obama | United States | 2008-2016 | Left | 0.15; 0.29 | Democratic Party |
| Donald Trump | United States | 2016-2020 | Right | 0.29 | Republican Party |

Note: [a] served a second term from 2014-2018.

**Table 4.** Summary Statistics & Correlation Matrix

|  | Mean | SD | Min | Max | (1) | (2) | (3) | (4) | (5) | (6) | (7) | (8) | (9) | (10) | (11) | (12) | (13) | (14) | (15) |
|---|---|---|---|---|---|---|---|---|---|---|---|---|---|---|---|---|---|---|---|
| *Individual-level* | | | | | | | | | | | | | | | | | | | |
| (1) OME | 0.059 | 0.24 | 0 | 1 | 1 | | | | | | | | | | | | | | |
| (2) Age | 42.3 | 14.03 | 14 | 99 | **-0.06** | 1 | | | | | | | | | | | | | |
| (3) Female | 0.51 | 0.50 | 0 | 1 | **-0.07** | **0.03** | 1 | | | | | | | | | | | | |
| (4) Secondary Education | 0.67 | 0.47 | 0 | 1 | **0.05** | **-0.12** | -0.00 | 1 | | | | | | | | | | | |
| (5) Household income tercile | 0.80 | 0.40 | 0 | 1 | **0.03** | **-0.04** | **-0.05** | **0.08** | 1 | | | | | | | | | | |
| (6) Fear of failure | 0.40 | 0.49 | 0 | 1 | **-0.09** | -0.03 | **0.07** | **-0.04** | -0.02 | 1 | | | | | | | | | |
| (7) Entrepreneurial self-efficacy | 0.48 | 0.50 | 0 | 1 | **0.21** | -0.00 | **-0.16** | **0.09** | **0.06** | **-0.14** | 1 | | | | | | | | |
| (8) Opportunity Recognition | 0.31 | 0.46 | 0 | 1 | **0.14** | **-0.07** | **-0.08** | **0.06** | **0.03** | **-0.08** | **0.16** | 1 | | | | | | | |
| *Country-level* | | | | | | | | | | | | | | | | | | | |
| (9) GDP per capita, PPP | 10.19 | 0.74 | 6.18 | 12 | **-0.03** | **0.11** | **0.02** | **0.09** | 0.00 | **0.02** | -0.01 | 0.01 | 1 | | | | | | |
| (10) Population (log) | 17.50 | 1.22 | 14.53 | 21 | **0.02** | **0.01** | -0.00 | -0.01 | **-0.05** | -0.00 | **0.02** | **-0.02** | **-0.21** | 1 | | | | | |
| (11) Populism | 0.17 | 0.22 | 0 | 1 | -0.01 | -0.01 | -0.00 | **0.03** | 0.01 | 0.01 | 0.01 | **-0.04** | **-0.11** | **-0.07** | 1 | | | | |
| (12) Checks and Balances | 1.70 | 0.32 | -0.06 | 2.06 | 0.00 | **0.06** | -0.00 | **-0.08** | **0.02** | **0.03** | **0.04** | **0.02** | **0.51** | **-0.06** | **-0.20** | 1 | | | |
| (13) Pro-Market Institutions | 7.48 | 0.52 | 5.76 | 8.42 | -0.00 | **0.13** | **0.03** | **0.10** | **-0.02** | **-0.04** | 0.01 | 0.01 | **0.68** | -0.03 | **-0.23** | **0.44** | 1 | | |
| (14) Left-Wing Ideology | 0.60 | 0.49 | 0 | 1 | -0.00 | **0.02** | **0.01** | **-0.14** | **-0.03** | 0.01 | **0.02** | **0.02** | **0.16** | **0.13** | **-0.23** | **0.32** | **0.17** | 1 | |
| (15) Right-Wing Ideology | 0.24 | 0.43 | 0 | 1 | -0.00 | -0.01 | -0.01 | **0.08** | **0.04** | 0.00 | -0.01 | **-0.04** | **-0.11** | -0.01 | **0.34** | **-0.17** | **-0.07** | **-0.69** | 1 |

Note: *Correlations for which p < 0.05 denoted in bold.*

**Table 5**: Main Results – Populism & Entrepreneurship

| | Opportunity-Motivated Entrepreneurship (OME) | | | |
|---|---|---|---|---|
| | (1) | (2) | (3) | (4) |
| *Explanatory Variables* | | | | |
| Populism | 0.980 | 1.008 | 2.097 | 1.409 |
| | (0.029) | (0.032) | (0.162) | (0.122) |
| Left-Wing Ideology | | 0.928 | 0.882 | 1.034 |
| | | (0.025) | (0.024) | (0.032) |
| Right-Wing Ideology | | 0.850 | 0.822 | 0.818 |
| | | (0.025) | (0.024) | (0.027) |
| Checks and balances | | 0.839 | 1.061 | 1.026 |
| | | (0.029) | (0.046) | (0.035) |
| *Interactions* | | | | |
| Checks and balances x Populism | | | 0.580 | |
| | | | (0.031) | |
| Left-Wing x Populism | | | | 0.433 |
| | | | | (0.058) |
| Right-Wing x Populism | | | | 0.975 |
| | | | | (0.127) |
| *Control Variables* | | | | |
| Age | 1.082 | 1.082 | 1.083 | 1.083 |
| | (0.002) | (0.003) | (0.003) | (0.003) |
| Age$^2$ | 0.999 | 0.999 | 0.999 | 0.999 |
| | (0.000) | (0.000) | (0.000) | (0.000) |
| Female | 0.726 | 0.726 | 0.726 | 0.726 |
| | (0.009) | (0.007) | (0.007) | (0.007) |
| Household income terciles | 1.407 | 1.409 | 1.409 | 1.407 |
| | (0.010) | (0.014) | (0.014) | (0.014) |
| Secondary education | 1.309 | 1.312 | 1.308 | 1.301 |
| | (0.022) | (0.029) | (0.029) | (0.028) |
| GDP per capita, ppp (log) | 0.903 | 0.985 | 0.779 | 0.870 |
| | (0.041) | (0.043) | (0.030) | (0.037) |
| Population (log) | 0.916 | 0.904 | 0.883 | 0.942 |
| | (0.014) | (0.013) | (0.013) | (0.013) |
| Fear of failure | 0.613 | 0.612 | 0.612 | 0.612 |
| | (0.007) | (0.007) | (0.007) | (0.007) |
| Entrepreneurial self-efficacy | 5.813 | 5.809 | 5.781 | 5.796 |
| | (0.014) | (0.079) | (0.078) | (0.079) |
| Opportunity recognition | 2.078 | 2.079 | 2.081 | 2.076 |
| | (0.021) | (0.020) | (0.020) | (0.020) |
| Christian | 1.022 | 1.048 | 1.095 | 1.081 |
| | (0.008) | (0.008) | (0.008) | (0.008) |
| Hindu | 1.033 | 1.068 | 1.104 | 1.084 |
| | (0.008) | (0.009) | (0.009) | (0.009) |
| Muslim | 1.056 | 1.047 | 1.088 | 1.068 |
| | (0.007) | (0.008) | (0.008) | (0.008) |
| Unaffiliated | 1.015 | 1.053 | 1.101 | 1.076 |
| | (0.007) | (0.007) | (0.008) | (0.008) |
| Buddhists | 1.017 | 1.072 | 1.098 | 1.093 |
| | (0.007) | (0.008) | (0.009) | (0.008) |
| Individualism | 1.000 | 1.002 | 0.998 | 0.999 |
| | (0.001) | (0.001) | (0.001) | (0.001) |
| Power distance | 1.003 | 1.003 | 1.011 | 0.994 |
| | (0.001) | (0.001) | (0.001) | (0.001) |
| Masculinity | 1.002 | 0.999 | 0.993 | 0.995 |

|                                   | Opportunity-Motivated Entrepreneurship (OME) | | | |
|---|---|---|---|---|
|                                   | (1) | (2) | (3) | (4) |
|                                   | (0.001) | (0.001) | (0.001) | (0.001) |
| Uncertainty                       | 1.012 | 1.012 | 1.004 | 1.006 |
|                                   | (0.002) | (0.001) | (0.001) | (0.001) |
| Long term                         | 1.007 | 1.006 | 1.009 | 1.009 |
|                                   | (0.002) | (0.002) | (0.002) | (0.002) |
| Indulgence                        | 1.009 | 1.012 | 1.017 | 1.005 |
|                                   | (0.002) | (0.002) | (0.002) | (0.002) |
| Harmony                           | 0.623 | 0.275 | 0.495 | 0.820 |
|                                   | (0.116) | (0.031) | (0.058) | (0.102) |
| Embedded                          | 0.394 | 0.214 | 0.404 | 0.196 |
|                                   | (0.089) | (0.022) | (0.039) | (0.021) |
| Hierarchy                         | 0.848 | 0.389 | 0.578 | 1.512 |
|                                   | (0.072) | (0.029) | (0.041) | (0.137) |
| Mastery                           | 0.590 | 0.863 | 1.323 | 0.934 |
|                                   | (0.114) | (0.099) | (0.162) | (0.110) |
| Affective Autonomy                | 0.972 | 0.653 | 0.968 | 0.800 |
|                                   | (0.085) | (0.050) | (0.075) | (0.060) |
| Intellectual Autonomy             | 0.363 | 0.240 | 0.477 | 0.290 |
|                                   | (0.072) | (0.019) | (0.036) | (0.023) |
| WVS trust                         | 0.991 | 0.990 | 0.999 | 0.996 |
|                                   | (0.001) | (0.001) | (0.001) | (0.001) |
| Standard deviation of inflation   | 1.025 | 1.005 | 1.015 | 0.984 |
|                                   | (0.009) | (0.011) | (0.011) | (0.012) |
| Standard deviation of tariff rates| 0.964 | 0.941 | 0.931 | 0.935 |
|                                   | (0.006) | (0.007) | (0.007) | (0.007) |
| Relative Political Extraction     | 1.019 | 1.021 | 1.020 | 1.013 |
|                                   | (0.003) | (0.003) | (0.003) | (0.003) |
| Tenure                            | 0.995 | 1.003 | 1.002 | 1.016 |
|                                   | (0.004) | (0.005) | (0.005) | (0.005) |
| EU member                         | 0.687 | 0.702 | 0.646 | 0.779 |
|                                   | (0.039) | (0.028) | (0.026) | (0.030) |
| Pro Market Institutions           | 1.429 | 1.426 | 1.366 | 1.396 |
|                                   | (.023) | (0.034) | (0.032) | (0.035) |
| Random part estimates             | | | | |
| Number of observations            | 780,280 | 780,280 | 780,280 | 780,280 |
| Number of groups (countries)      | 33 | 33 | 33 | 33 |
| Variance of random intercept      | 0.097 | 0.106 | 0.079 | 0.078 |
| % of variance, ρ                  | 2.09 | 3.12 | 2.35 | 2.32 |
| Model fit statistics              | | | | |
| Degrees of freedom                | 51 | 54 | 55 | 56 |
| Prob > $\chi^2$                   | 0.000 | 0.000 | 0.000 | 0.000 |
| Log-likelihood                    | -164,313 | -164,278 | -164,245 | -164,246 |
| Year dummies?                     | Yes | Yes | Yes | Yes |
| AIC [a]                           | 328,820 | 328,663 | 328,600 | 328,605 |
| LR test of ρ = 0 [b]              | *** | *** | *** | *** |
| LR test of model fit [c]          | -- | -- | *** | *** |

*Note*: Standard errors in parentheses. Estimates reported as odds ratios (OR). OR > 1 denotes a positive relationship, and OR < 1 denotes a negative relationship. [a] AIC is Akaike's information criterion = 2k - 2(log-likelihood), where k denotes the degrees of freedom (number of predictors in the model). [b] Statistically significant ($p < 0.001$). LR test of ρ = 0 confirms that the country-level variance component is important. This is the ICC. [c] LR test performed between Model 2 and either Model 3, or Model 4. Models estimated using multilevel logit regression.

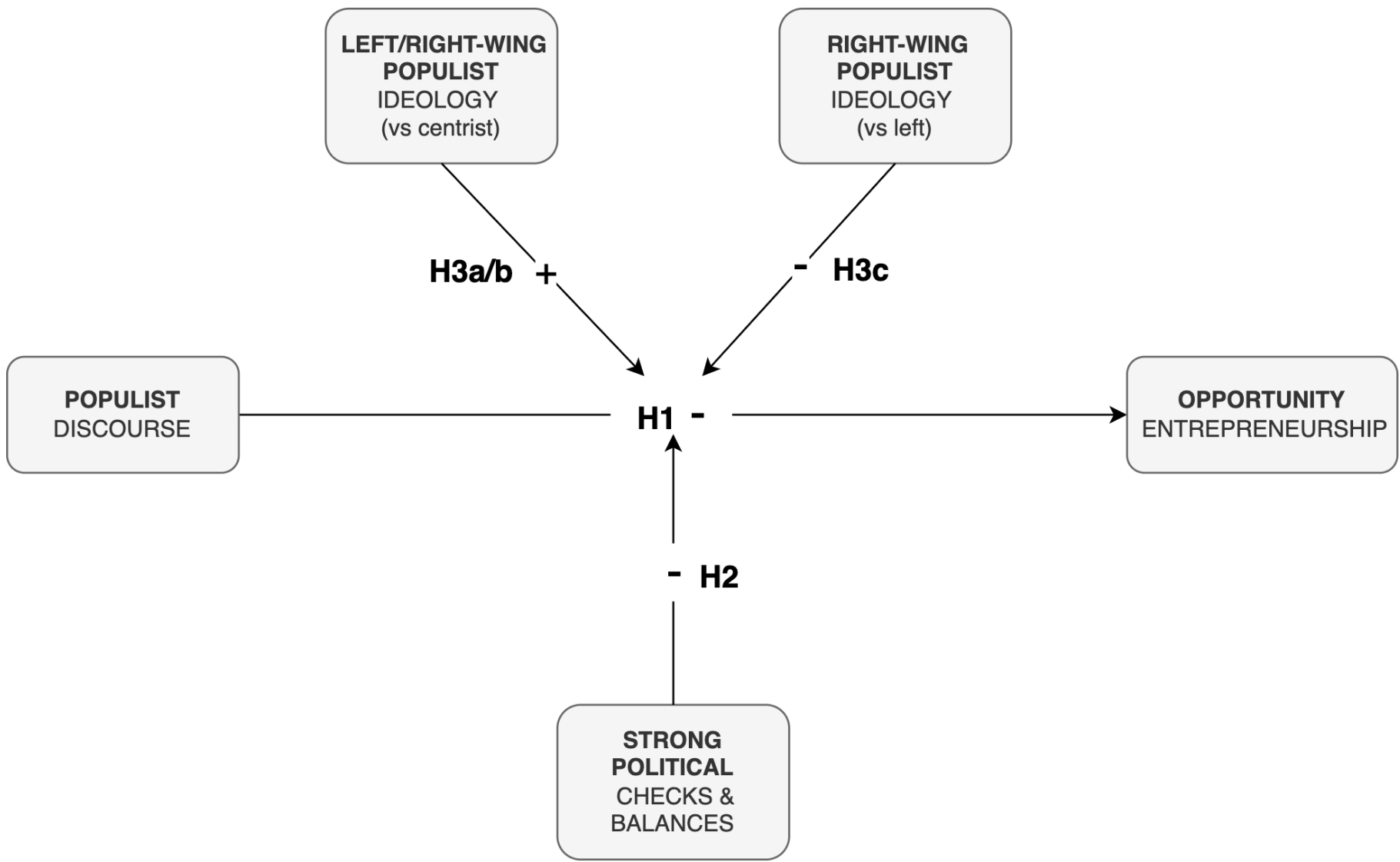

**Figure 1:** Theoretical Model

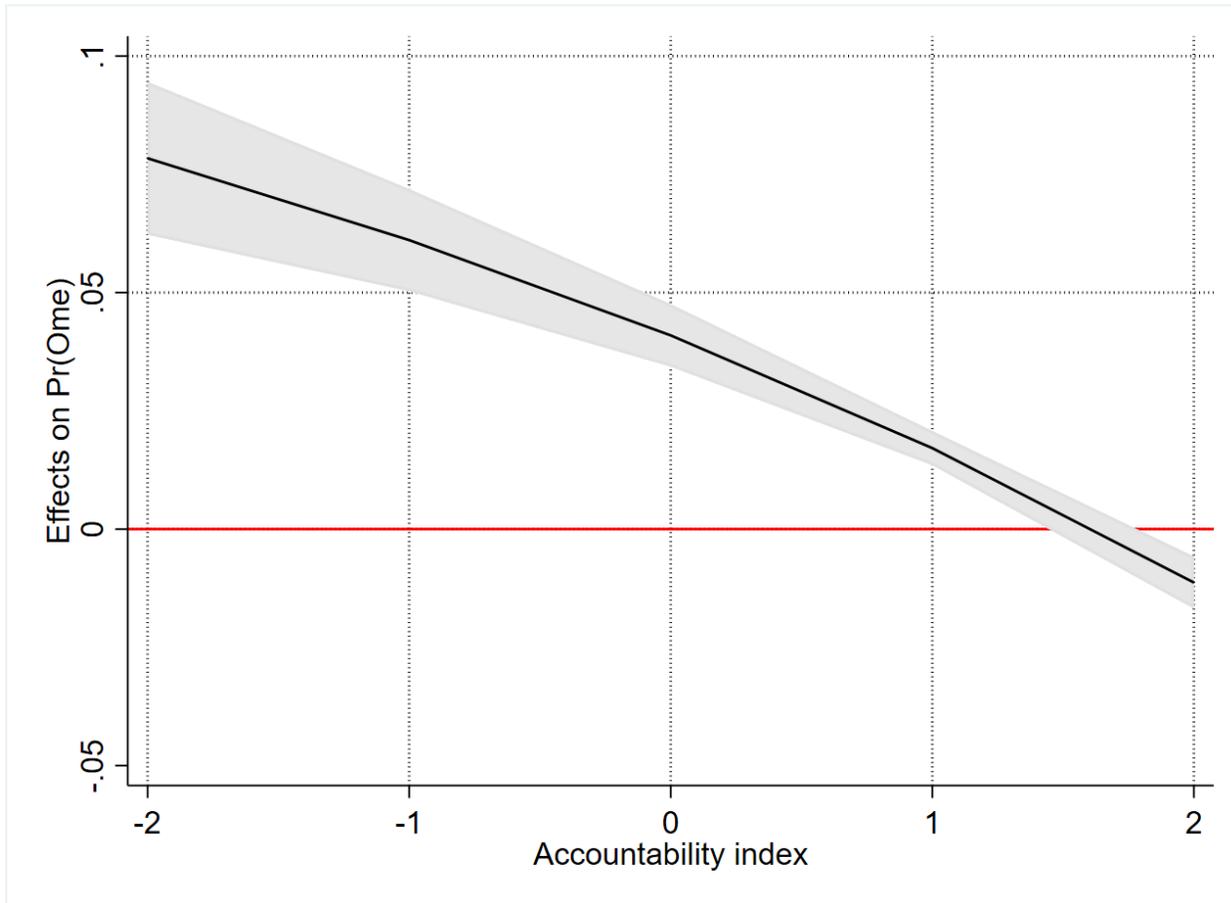

**Figure 2.** Checks and balances moderates the relationship between populism and OME

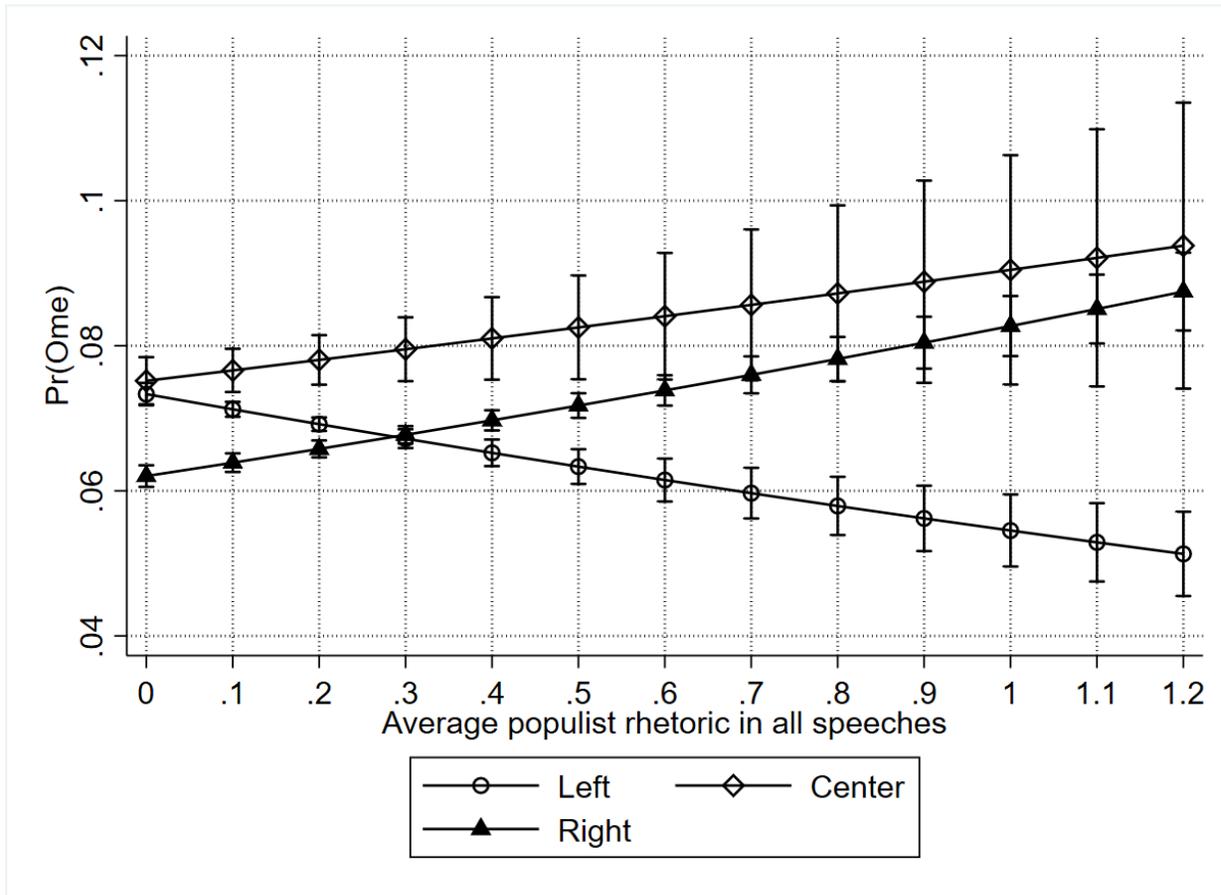

**Figure 3.** Moderating results for centrist, left-wing, and right-wing ideologies.

## ACKNOWLEDGEMENTS

We thank three anonymous referees for providing constructive and development feedback throughout the review process and Editor Arjen van Witteloostuijn for providing excellent and encouraging guidance. We also appreciate comments and suggestions received by participants at the University of Louisville College of Business Research Colloquium and the Academy of Management and Babson College Entrepreneurship Research Conferences.

## AUTHOR BIOGRAPHIES

**Daniel L. Bennett** is an Assistant Professor of Entrepreneurship and Assistant Director of the Center for Free Enterprise at the University of Louisville. His research examines how the institutional and policy context influences entrepreneurship and innovation, and the implications for economic development. He obtained his Ph.D. from Florida State University.

**Christopher Boudreaux** is an Associate Professor of Economics at Florida Atlantic University and a Research Fellow for the Phil Smith Center for Free Enterprise. His research interests include entrepreneurship, innovation, and the economic analysis of public policy. He earned his PhD from Florida State University and taught at Texas A&M International University prior to FAU.

**Boris Nikolaev** is an Assistant Professor of Entrepreneurship at Baylor University. His research interests include public policy, applied microeconomics, entrepreneurship, economic development, new institutional economics, and mental health and well-being. He received his Ph.D. in Economics from the University of South Florida. He will be joining the department of Management at Colorado State University in the summer of 2022.

## ENDNOTES

[1] Weyland (2017, p. 53) offers two additional potential limitations of the ideation approach that we consider secondary concerns because they largely reflect a difference of opinion in the key defining characteristics of populism: (1) Hawkin's methodological innovation that enabled the measurement of populism as discourse using the holistic grading approach has the potential to produce "important `false positives'"; and (2) the ideation approach misunderstands populism's intention and distorts its meaning, which according to Weyland, "revolves around top-down leadership."

[2] Given these results, we followed Stócki and Rode (2021) by re-estimating the populism-political ideology interaction model treating centrist and right-wing ideology as the baseline. The results suggest that left-wing populist discourse strengthens the negative relationship between populist discourse and OME, relative to populist discourse by either a centrist or right-wing leader.